\documentclass[11pt,a4paper,reqno]{amsart}
\usepackage{cite}
\usepackage{multirow}
\usepackage[centertags]{amsmath}
\usepackage{amsfonts}
\usepackage{amssymb}
\usepackage{amsthm}
\usepackage{newlfont}
\usepackage[pdftex]{graphicx,color}
\usepackage{amscd}
\usepackage{verbatim}
\allowdisplaybreaks
\usepackage{multirow}

\newtheorem{Theorem}{Theorem}[section]
\newtheorem{Proposition}{Proposition}

\theoremstyle{definition}
\newtheorem{Definition}{Definition}
\newtheorem{Example}{Example}
\newenvironment{Remark}{\medskip\noindent{\sc Remark.}}{\medskip}

\input xypic

\usepackage{lscape}
\usepackage{longtable}
\usepackage{amssymb}
\usepackage{epsfig}
\usepackage{graphics}
\usepackage{graphicx}
\usepackage{amsmath}
\usepackage{amsthm}
\usepackage{amscd}
\usepackage{verbatim}
\usepackage{xcolor}
\usepackage{pdflscape}
\usepackage{afterpage}
\usepackage{capt-of}
\usepackage{float}
\usepackage{tikz}
\usepackage{tikz-cd}
\usetikzlibrary{positioning}
\usetikzlibrary{shapes.misc}
\usepackage{xcolor}
\usepackage{hyperref}

\hypersetup{
     colorlinks   = true,
     linkcolor    =  {blue!55!green!90!white},
     citecolor    =  {blue!55!green!90!white}}
\tikzset{cross/.style={cross out, draw=black, minimum size=2*(#1-\pgflinewidth), inner sep=0pt, outer sep=0pt},cross/.default={1pt}}
\tikzstyle{block}=[draw, rectangle, text width=4em, text centered, minimum height = 4mm, node distance=3em]
\tikzstyle{line}=[draw, -stealth, thick]
\tikzstyle{Block}=[draw, rectangle, text width=6em, text centered, minimum height = 4mm, node distance=3em]

\definecolor{mycolor}{rgb}{0.21, 0.46, 0.53}

\def\R{\mathbb R}

\def\N{\mathbb N}

\def\a{\alpha}

\begin{document}

\title[Nested Polyhedra and Indices]{Nested Polyhedra and Indices of Orbits of Coxeter Groups of Non-Crystallographic Type}


\author[M.Myronova]{Mariia Myronova 
 $^{1}$}
 \author[J.Patera]{Ji\v{r}\'{\i} Patera $^{2}$}
 \author[M.Szajewska]{Marzena Szajewska $^{3}$}








\begin{abstract}The invariants of finite-dimensional representations of simple Lie algebras, such as even-degree indices and anomaly numbers, are considered in the context of the non-crystallographic finite reflection groups $H_2$, $H_3$ and $H_4$. Using a representation-orbit replacement, the definitions and properties of the indices are formulated for individual orbits of the examined groups. The indices of orders two and four of the tensor product of $k$ orbits are determined. Using the branching rules for the non-crystallographic Coxeter groups, the embedding index is defined similarly to the Dynkin index of a representation. Moreover, since the definition of the indices can be applied to any orbit of non-crystallographic type, the algorithm allowing to search for the orbits of smaller radii contained within any considered one is presented for the Coxeter groups $H_2$ and $H_3$. The geometrical structures of nested polytopes are exemplified.
\end{abstract}

\maketitle





{
\footnotesize
\noindent
$^{1}$ \quad D\'epartement de Physique, Universit\'e de Montr\'eal, Complexe des Sciences, 1375 Avenue Th\'er\`ese-Lavoie-Roux, Montr\'eal, QC H2V 0B3, Canada; maria.myronova@umontreal.ca

\noindent
$^{2}$ \quad Centre de recherches math\'ematique, Universit\'e de Montr\'eal, C.~P.~6128 Centre-Ville, Montr\'eal, QC H3C\,3J7, Canada; patera@crm.umontreal.ca

\noindent
$^{3}$ \quad Department of Mathematics, University of Bia{\l}ystok, 1M Cio{\l}kowskiego, PL-15-245 Bia{\l}ystok, Poland; \newline m.szajewska@math.uwb.edu.pl

}

\section{Introduction}

The purpose of this paper is to formulate the definitions of the even- and odd-degree indices for orbits of the non-crystallographic Coxeter groups $H_2$, $H_3$ and $H_4$ (the symmetry group of a regular pentagon, a regular icosahedron and the 600-cell, respectively). In this case, the definition of the indices of irreducible representations of simple Lie algebras provides a foundation for the indices of orbits. The generalization of properties of the formulated indices is achieved by examining the individual orbits of the investigated groups. 

A significant number of applications of non-crystallographic Coxeter groups in solid-state physics~\cite{SSP}, chemistry \cite{NSL} and structural genomics \cite{Ter} motivates the current study. The symmetries of the $H_2$ and $H_4$ groups play an essential role in the construction and description of quasicrystals \cite{LR}. The icosahedral symmetry of the Coxeter group $H_3$ reveals the structure of the extensive diversity of spherical molecules \cite{FM}. Moreover, during the past few years, the $H_3$ group has gained considerable interest in mathematical virology, since it serves as a blueprint for examining and describing the architecture and assembly of spherical viruses \cite{Dech, IGSRZT, SICKT, Tw1}.

The pertinent information about the non-crystallographic reflection groups $H_n$, $n\in\{2,3,4\}$ can be found in \cite{CMP, Shch}. Even though any Weyl orbit is linked to a finite-dimensional representation of a Lie algebra, this relation does not exist for the non-crystallographic cases. In general, any orbit $O_{\lambda}(G)$ of a finite Coxeter group $G$ arises from the action of the corresponding reflections on the dominant (seed) point $\lambda \in \mathbb{R}^{n}$. The coordinates of $\lambda$ are commonly provided in the $\omega-$basis, and they take values of any non-negative real numbers. Since any orbit of $H_3$ can be represented geometrically by a Euclidean (spherical) polytope, the variation of the coordinates of $\lambda\in\R^3$ impacts the lower-dimensional faces represented by edges (arcs) and polygons (tiles on a sphere).  As a result, a deeper understanding of a chosen seed point is achieved. The numbers and types of faces of a polytope are determined using the decoration procedure applied to a Coxeter--Dynkin diagram \cite{CKPS}.

During the past decades, it has been convenient to characterize representations of simple Lie algebras by their dimensions \cite{Kir, Ram}. Even though the formula for the dimension is well-known, its~difficulty in practical exploitation rapidly increases together with the rank of a corresponding Lie algebra. To overcome this problem, E.B. Dynkin introduced the index that can be calculated for any irreducible representation \cite{Brad, Pany}. Since then, the ``Dynkin index'' is considered as a powerful tool for the classification of semisimple subalgebras of simple Lie algebras \cite{Dyn}. The further development of research led to the discovery of the higher-degree indices of finite irreducible representations that have been formulated in \cite{PSW}. Since the decomposition of the tensor product of representations of a simple Lie algebra into the direct sum of irreducible components is important and relevant to many branches of physics, the general formulas for indices of such decompositions are determined in \cite{OP1, OP2}. 

In this paper, we define the analogs of the higher-degree indices replacing irreducible representations of simple Lie algebras by orbits of the non-crystallographic finite reflection groups. This approach yields several advantages, since the orbit size is finite, and the product of orbits is always decomposable. Hence, the even- and odd-degree indices of orbits of non-crystallographic type are formulated in Sections~\ref{evensec} and~\ref{oddsec}. The former include the lower-order indices of the tensor product of orbits. The latter are recognized in physics as the anomaly numbers, since they determine the symmetry-breaking parameters defined for particle systems \cite{Okubo1, OP3, PS}. For the odd-degree index, it~is only necessary to determine whether it is zero or not. In our framework, such indices are considered as a generalization of the anomaly numbers of irreducible representations \cite{ZOT}.

The Dynkin index remains a valid invariant only if a single orbit of non-crystallographic type is involved in its definition. Therefore, in Section~\ref{embsec}, we explore the analog of such an invariant called the embedding index. The calculations of the index proceed whenever the branching rule for a finite reflection group and its subgroup is known ($G' \subset G$). For crystallographic reflection groups, the branching rules are determined for the rank up to $n=8$ (for instance, see \cite{LNP, LP2} and references therein). Recently, the branching rules have been formulated for the non-crystallographic reflection groups as well \cite{GPS}. 

Furthermore, since we are restricted to the non-crystallographic groups, we introduce a search algorithm to find the orbits of smaller radii that may appear inside an initial one (Section~\ref{detsec}). Here,~such orbits are referred to as `lower orbits.' The subtraction of the simple roots of $H_n$, $n\in\{2,3,4\}$ from a seed point $\lambda$ with its coordinates in the $\omega-$basis provides the dominant points of lower orbits. We~demonstrate that this method coincides with the root-subtraction for orbits of crystallographic type. Such a procedure forms a weight system similar to the one of representation. In the geometrical interpretation, any obtained set of lower orbits together with a starting one results in the structure of nested polyhedra \cite{Jan, TTVZ, Zel}. Such a set of polytopes is quite unusual, as it differs from the sets obtained for crystallographic cases. For the latter, whenever any two polytopes with dominant points consecutively obtained by the subtraction method are considered, one can notice that each vertex of a larger polytope is found in the middle of each edge of a polytope of smaller radius. In contrast, the nested polyhedra of non-crystallographic types do not have this property.
 
\section{Even-Degree Indices for Orbits}\label{evensec}

The important information about the even-degree indices of representations of simple Lie algebras can be found in several papers \cite{OP1, Ram, PSW}. In this section, considered analogs possess the same properties as the decomposition of products does. However, this property is limited to the indices of degrees two, four and, for some groups, six. Replacing an irreducible representation of a simple Lie algebra with an orbit of a finite reflection group has several advantages:
\begin{itemize}
\item the size of an orbit of any Coxeter group is always limited;
\item the points of an orbit have only real numbers as their coordinates;
\item the product of several orbits can always be decomposed into the sum of orbits of smaller sizes.
\end{itemize}

\begin{Definition}
Let $G$ be a finite reflection group, and $O_\lambda(G)$ be an orbit of elements with a dominant point $\lambda$. The number defined by
\begin{equation*}
I_\lambda^{2p}(G)=\sum_{\mu\in O_\lambda(G)} \langle \mu, \mu \rangle^{p}, \quad p\in\mathbb{N},
\end{equation*}
is called the {$2p$-order index} of an orbit $O_\lambda(G)$. The summation extends over all the elements of $O_\lambda(G)$, and~$\langle\cdot,\cdot\rangle$ denotes the scalar product in the weight space of $G$.
\end{Definition}
\begin{Remark}
For $p=0$, the zero order index of an orbit $O_\lambda(G)$ is equal to its size:
\begin{equation*}
I_\lambda^{0}(G)=|O_\lambda(G)|,
\end{equation*}
where $|O_\lambda(G)|$ denotes the size of an orbit generated by the action of $G$ on a seed point $\lambda$. The sizes of orbits of the examined non-crystallographic groups are presented in Table~\ref{orbit}.
\end{Remark}

Since the elements of any orbit $O_\lambda(G)$ are equidistant from the origin, we have the following~remark.

\begin{Remark}
The formula for even-degree indices has the form:
\begin{equation}
I_\lambda^{2p}(G)=|O_\lambda(G)|\langle \lambda, \lambda\rangle^ p, \quad p\in\mathbb{N}.
\label{indnc}
\end{equation}
\end{Remark}

\begin{Proposition}\label{indH2}
For the non-crystallographic reflection groups, the general formulas for $2p$-order indices are the following ones:
\begin{align*}
(3{-}\tau)^pI_{(a_1,a_2)}^{2p}(H_2)=& |O_{(a_1,a_2)}(H_2)|\cdot [2(a_1^2{+}\tau a_1 a_2{+}a_2^2)]^p,\\
(4{-}2\tau)^pI_{(a_1,a_2,a_3)}^{2p}(H_3)=&|O_{(a_1,a_2,a_3)}(H_3)| \cdot [(3{-}\tau)a_1^2{+}4a_2^2{+}3a_3^2{+}4 a_1 a_2{+}2\tau a_1 a_3{+}4\tau a_2 a_3]^p,\\
(5-3\tau)^pI_{(a_1,a_2,a_3,a_4)}^{2p}(H_4)=&|O_{(a_1,a_2,a_3,a_4)}(H_4)|\cdot [2((2{-}\tau)a_1^2{+}(3{-}\tau)a_2^2{+}3a_3^2{+}2a_4^2{+}(3{-}\tau)a_1a_2\\
&{+}2a_1a_3{+}\tau a_1a_4{+}4 a_2a_3{+}2\tau a_2a_4{+}3\tau a_3a_4)]^p,
\end{align*}
where $\tau=\frac{1+\sqrt{5}}{2}=1.618\ldots$ is the positive solution of the quadratic equation $x^2=x+1$ known as the golden~ratio.
\end{Proposition}

\begin{proof}
The inner product $\langle\cdot,\cdot\rangle$ of the elements of orbits of the non-crystallographic groups $H_n$ has the following form:
\begin{equation}\label{inprod}\langle (a_1,\ldots,a_n), (b_1,\ldots,b_n) \rangle = \left(\begin{matrix}a_1 \ldots  a_n \end{matrix}\right) C^{-1}_{H_n}  \left(\begin{matrix}b_1 \\\vdots \\ b_n \end{matrix}\right), \quad n\in\{2,3,4\}.
\end{equation}
where $C^{-1}_{H_n}$ is the inverse Cartan matrix (Table~\ref{Cartan}).
Applying \eqref{inprod} to \eqref{indnc}, the desired formulas can be immediately obtained.
\end{proof}

\begin{table}[H]
\caption{The sizes of  orbits $O_\lambda(H_n)$ of the non-crystallographic groups $H_n$, $n\in\{2,3,4\}$ provided for each type of a dominant point $\lambda$ with the coefficients $a,b,c,d \in \R^{>0}$.}\label{orbit}
\centering
\begin{tabular}{ccc}
\scalebox{0.9}[0.9]{\begin{tabular}{c|c}
\boldmath{$\lambda$} & \boldmath{$|O_\lambda(H_2)|$}\\
\hline
$(a,0)$ & 5\\
$(0,b)$ & 5\\
$(a,b)$ & 10
\bigskip
\bigskip
\bigskip
\bigskip
\medskip
\medskip
\\ 
\boldmath{$\lambda$} & \boldmath{$|O_\lambda(H_3)|$}\\
\hline
$(a,0,0)$ & 12\\
$(0,b,0)$ & 30\\
$(0,0,c)$ & 20 \\
$(a,b,0)$ & 60 \\
$(a,0,c)$ & 60\\
$(0,b,c)$ & 60 \\
$(a,b,c)$ & 120\\
\end{tabular} \qquad
\begin{tabular}{c|c}
\boldmath{$\lambda$} & \boldmath{$|O_\lambda(H_4)|$}\\
\hline
$(a,0,0,0)$ & 120 \\
$(0,b,0,0)$ & 720\\
$(0,0,c,0)$ &  1200\\
$(0,0,0,d)$ &  600\\
$(a,b,0,0)$ & 1440 \\
%
$(a,0,c,0)$ & 3600\\
$(a,0,0,d)$ &  2400\\
$(0,b,c,0)$ &  3600\\
$(0,b,0,d)$ &  3600\\
$(0,0,c,d)$ &  2400\\
$(a,b,c,0)$ &7200 \\
$(a,b,0,d)$ & 7200\\
$(a,0,c,d)$ & 7200\\
$(0,b,c,d)$ & 7200\\
$(a,b,c,d)$ &14,400\\
\end{tabular}}
\end{tabular}
\medskip
\end{table}
\unskip

\begin{table}[H]
\caption{The Cartan matrices and their inverses for the non-crystallographic groups $H_2$, $H_3$ and $H_4$.}\label{Cartan}
\centering
\begin{tabular}{ll}
$C_{H_2}=\left(\begin{array}{cc}2&-\tau\\ -\tau&2 \end{array}\right)$ &
$C^{-1}_{H_2}=\frac1{3-\tau}\left(\begin{array}{cc}2&\tau\\ \tau&2\end{array}\right)$\\\\
$C_{H_3}=\left(\begin{array}{ccc}2&-1&0 \\ -1&2&-\tau\\ 0&-\tau&2\end{array}\right)$ &
$C^{-1}_{H_3}=\frac12\left(\begin{array}{ccc}2+\tau&2+2\tau&1+2\tau \\ 2+2\tau&4+4\tau&2+4\tau\\ 1+2\tau&2+4\tau&3+3\tau\end{array}\right)$\\\\
$C_{H_4}=\left(\begin{array}{cccc} 2&-1&0&0 \\ -1&2&-1&0\\ 0&-1&2&-\tau\\0&0&-\tau&2\end{array}\right)$&
$C^{-1}_{H_4}=\left(\begin{array}{cccc} 2+2\tau&3+4\tau&4+6\tau&3+5\tau \\ 3+4\tau&6+8\tau&8+12\tau&6+10\tau\\ 4+6\tau&8+12\tau&12+18\tau&9+15\tau\\3+5\tau&6+10\tau&9+15\tau&8+12\tau\end{array}\right)$ \\
\end{tabular}
\end{table}

\begin{Definition}
Let $G$ be a finite reflection group. The direct sum of orbits with dominant points $\lambda_1, \ldots,\lambda_k$, $k \geq 2$, is provided by the formula:
\begin{equation}\label{dirsum}
O_{\lambda_1\oplus \ldots\oplus \lambda_k}(G) = \bigcup_{\mu_i \in O_{\lambda_i(G)}\atop i\in\{1,2,\ldots,k\}} \mu_i=O_{\lambda_1}(G)\cup \ldots\cup O_{\lambda_k}(G).
\end{equation}
The size of the direct sum is equal to
\begin{equation}\label{dirsumsize}
|O_{\lambda_1\oplus\ldots\oplus \lambda_k}(G)| =|O_{\lambda_1}(G)|+\ldots+|O_{\lambda_k}(G)|.
\end{equation}
\end{Definition}

\begin{Definition}
Let $G$ be a finite reflection group. The tensor product of orbits of $G$ with dominant points $\lambda_1, \ldots,\lambda_k$, $k \geq 2$, is provided by the summation of elements of each orbit with elements of other orbits as
\begin{equation}\label{tenprod}
O_{\lambda_1\otimes \ldots\otimes \lambda_k}(G) = \bigcup_{\mu_i \in O_{\lambda_i(G)}\atop i\in\{1,2,\ldots,k\}} (\mu_1+\ldots+\mu_k).
\end{equation}

The size of the tensor product is equal to
\begin{equation}\label{tenprodsize}
|O_{\lambda_1\otimes\ldots\otimes \lambda_k}(G)| =|O_{\lambda_1}(G)|\cdot\ldots\cdot|O_{\lambda_k}(G)|.
\end{equation}
\end{Definition}

\begin{Remark}
The tensor product of $k$ orbits of $G$ decomposes into a union of several orbits \cite{HLP}. In this case, the~highest weight is $\lambda_1+\ldots+\lambda_k$, and  the product of orbits decomposes as follows:
\begin{equation*}
{\lambda_1\otimes \ldots\otimes \lambda_k} = (\lambda_1+\ldots+\lambda_k)\cup \ldots \cup \textrm{ other lower-order orbits}.
\end{equation*}
\end{Remark}

\begin{Example}
Let us consider two orbits, $O_{(1,0)}(H_2)$ and $O_{(0,\tau)}(H_2)$. The direct sum (\ref{dirsum}) and tensor product (\ref{tenprod}) of orbits are written explicitly as
\begin{align*}
O_{(1,0)\oplus(0,\tau)}(H_2)=&\{(1,0),({-}1,\tau),(\tau,{-}\tau),({-}\tau,1),(0,{-}1),\\
& (0,\tau),(\tau+1,{-}\tau),({-}\tau{-}1,\tau+1),(\tau,{-}\tau{-}1),({-}\tau,0)\};\\
O_{(1,0)\otimes(0,\tau)}(H_2)=&\{(1,\tau ),(\tau {+}2,{-}\tau ),({-}\tau ,\tau {+}1),(\tau {+}1,{-}\tau {-}1),(1{-}\tau ,0),({-}1,2 \tau ),\\
&(\tau ,0),({-}\tau {-}2,2 \tau {+}1),(\tau {-}1,{-}1),({-}\tau {-}1,\tau ),(\tau ,0),(2 \tau {+}1,{-}2 \tau ),\\
&({-}1,1),(2 \tau ,{-}2 \tau {-}1),(0,{-}\tau ),({-}\tau ,\tau {+}1),(1,1{-}\tau ),({-}2 \tau {-}1,\tau {+}2),\\
&(0,{-}\tau ),({-}2 \tau ,1),(0,\tau {-}1),(\tau {+}1,{-}\tau {-}1),({-}\tau {-}1,\tau ),(\tau ,{-}\tau {-}2),({-}\tau ,{-}1)\}.
 \end{align*}
The tensor product of two orbits decomposes into the union of orbits as
\begin{equation*}\mathop{(1,0)}_{5}\mathop{\otimes}_{\cdot}\mathop{(0,\tau)}_{5}\mathop{=}_{=} \mathop{(1,\tau)}_{10}\mathop{\cup}_{+}\mathop{2(\tau,0)}_{2\cdot 5}\mathop{\cup}_{+}\mathop{(0,\tau{-}1)}_{5}.
\end{equation*}

The numbers attached to the dominant points correspond to the sizes of the orbits of $H_2$ provided by (\ref{dirsumsize}) and (\ref{tenprodsize}) (see Table~\ref{orbit}). The number of elements of the orbit product is equal to the number of elements after the decomposition.
\end{Example}

\begin{Proposition}\label{prop1}
Let $G$ be a finite reflection group. The formulas for lower-order indices of the tensor product of $k$ orbits of $G$ are given by:
\begin{eqnarray}\label{1}
(i)&I^{2}_{\lambda_1{\otimes}{\cdots} {\otimes} \lambda_k}(G)=\prod\limits_{i=1}^k I^{0}_{\lambda_i}(G) \sum\limits_{j=1}^k \frac{I^{2}_{\lambda_j}(G)}{|O_{\lambda_j}(G)|} =\sum\limits_{j=1}^k \left(  I^{2}_{\lambda_j}(G) \prod\limits_{i\ne j\atop i=1}^k I^{0}_{\lambda_i}(G)\right), \nonumber\\
(ii)&I^{4}_{\lambda_1{\otimes}{\cdots} {\otimes} \lambda_k}(G) =\sum\limits_{j=1}^k \left( I^{4}_{\lambda_j}(G) \prod\limits_{i\ne j \atop i=1}^k I^{0}_{\lambda_i}(G)\right)+\frac{2 (r+2)}{r} \sum\limits_{j,l=1\atop j\ne l}^k \left( I^{2}_{\lambda_j}(G)I^{2}_{\lambda_l}(G) \prod\limits_{{i\ne j,l}\atop{ i=1}}^k I^{0}_{\lambda_i}(G)\right), \nonumber
\end{eqnarray}
where $k\in \mathbb{N}^{\geq2}$, and $r$ denotes the rank of $G$. 
\end{Proposition}

\begin{proof}
$(i)$ Using the Equation \eqref{indnc}, we immediately have
\begin{align*}
I^{2}_{\lambda_1{\otimes}{\cdots} {\otimes} \lambda_k}(G)&=|O_{\lambda_1{\otimes}{\cdots} {\otimes} \lambda_k}(G)|\cdot \langle \lambda_1{\otimes}{\cdots} {\otimes} \lambda_k,\lambda_1{\otimes}{\cdots} {\otimes} \lambda_k \rangle \\
&= |O_{\lambda_1}(G)|\cdot|O_{\lambda_2}(G)|\ldots|O_{\lambda_k}(G)|\cdot\left( \langle \lambda_1,\lambda_1 \rangle +\langle \lambda_2,\lambda_2 \rangle+\ldots+\langle \lambda_k,\lambda_k \rangle\right) \\
&= \prod\limits_{ i=1}^k I^{0}_{\lambda_i}(G)\cdot\sum\limits_{j=1}^k  \langle \lambda_j,\lambda_j \rangle=\sum\limits_{j=1}^k \left(  I^{2}_{\lambda_j}(G) \prod\limits_{i\ne j\atop i=1}^k I^{0}_{\lambda_i}(G)\right).
\end{align*}

$(ii)$ Let us recall the pertinent properties of orbits of finite reflection groups. Considering any point $\mu=(\mu_1,\ldots,\mu_r)$, where $r=rank\ G$, we obtain
\begin{align*}
\sum_{\mu\in O_\lambda(G)} \mu_i &=0, \qquad \sum_{\mu\in O_\lambda(G)} \mu_i\mu_j =\frac{\delta_{ij}}{r}\sum_{\mu\in O_\lambda(G)} \mu^2.
\end{align*}
Hence, the index for $p=2$ can be written as
\begin{align*}
I^{4}_{\lambda_1{\otimes}{\cdots} {\otimes} \lambda_k}(G)&=|O_{\lambda_1{\otimes}{\cdots} {\otimes} \lambda_k}(G)|\cdot \langle \lambda_1{\otimes}{\cdots} {\otimes} \lambda_k,\lambda_1{\otimes}{\cdots} {\otimes} \lambda_k \rangle^2 \\
&= |O_{\lambda_1}(G)|\cdot|O_{\lambda_2}(G)|\ldots|O_{\lambda_k}(G)|\cdot\left(\sum_{i,j=1 }^k\langle \lambda_i,\lambda_i \rangle\langle \lambda_j,\lambda_j \rangle+4\sum_{i,j=1}^k\langle \lambda_i,\lambda_j \rangle^2\right).
\end{align*}

Using the properties of orbits, we obtain the following expressions:
\begin{align*}
& |O_{\lambda_1}(G)|\cdot|O_{\lambda_2}(G)|\ldots|O_{\lambda_k}(G)|\cdot\left( \sum_{i=1}^k\langle \lambda_i,\lambda_i \rangle^2+2 \sum_{i,j=1 \atop i\ne j }^k\langle \lambda_i,\lambda_i \rangle\langle \lambda_j,\lambda_j \rangle+\frac{4\delta_{ij}}{r}\sum_{i,j=1}^k\langle \lambda_i,\lambda_j \rangle^2\right)\\
&= |O_{\lambda_1}(G)|\cdot|O_{\lambda_2}(G)|\ldots|O_{\lambda_k}(G)|\cdot\left( \sum_{i=1}^k\langle \lambda_i,\lambda_i \rangle^2+\frac{2(r+2)}{r} \sum_{i,j=1 \atop i\ne j }^k\langle \lambda_i,\lambda_i \rangle\langle \lambda_j,\lambda_j \rangle\right)\\
&= \sum\limits_{j=1}^k \left( I^{4}_{\lambda_j}(G) \prod\limits_{i\ne j \atop i=1}^k I^{0}_{\lambda_i}(G)\right)+\frac{2 (r+2)}{r} \sum\limits_{j,l=1\atop j\ne l}^k \left( I^{2}_{\lambda_j}(G)I^{2}_{\lambda_l}(G) \prod\limits_{{i\ne j,l}\atop{ i=1}}^k I^{0}_{\lambda_i}(G)\right).
\end{align*}
\end{proof}

\begin{Remark}
In general, the indices of $k$-th tensor product of orbits of a group $G$ are defined recursively as
\begin{equation*}
I^{2p}_{\lambda_1{\otimes}{\cdots} {\otimes} \lambda_k}(G) = I^{2p}_{\lambda_1{\otimes} (\lambda_2\otimes{\cdots} {\otimes} \lambda_k)}(G) ,\quad  k\in \mathbb{N}^{\geq2}.
\end{equation*}
\end{Remark}

The obvious observation is 
\begin{equation*}
I^{2p}_{{\lambda_1\oplus \ldots\oplus \lambda_k}}(G)=\sum_{i=1}^k I^{2p}_{\lambda_i}(G).
\end{equation*}

\begin{Example}
Let us calculate the second-order index of the tensor product of the orbits $O_{(1,0)}(H_2)$ and $O_{(0,1)}(H_2)$. Such a product decomposes as
\begin{equation*}
(1,0)\otimes (0,1)=(1,1) \cup (\tau-1,\tau-1) \cup 2(0,0).
\end{equation*}

Therefore, using the decomposition given above, we can calculate the second-degree index as follows: 
\begin{equation*}
I^2_{(1,0)\otimes (0,1)}(H_2) = I^2_{(1,1)}(H_2)+ I^2_{(\tau-1,\tau-1)}(H_2)+2 I^2_{(0,0)}(H_2)=\frac{20(\tau+2)}{3-\tau}+20+0=\frac{100}{3-\tau}.
\end{equation*}

Taking into consideration Proposition~\ref{prop1}, the same result is obtained:
\begin{equation*}
I^2_{(1,0)\otimes (0,1)}(H_2) = I^2_{(1,0)}(H_2)I^0_{(0,1)}(H_2)+ I^0_{(1,0)}(H_2)I^2_{(0,1)}(H_2)=5\cdot\frac{10}{3-\tau}+5\cdot\frac{10}{3-\tau}=\frac{100}{3-\tau}.
\end{equation*}
\end{Example}

\begin{Proposition}
Let $G=G_1 {\times}\ldots{\times} G_k$ be a finite reflection group. 
The formula for $2p$-order indices of the product of $k$ orbits $\lambda_i\in O(G_i)$, $i\in\{1,2,\ldots,k\}$ is provided by:
\begin{eqnarray}\label{prodsum}
I^{2p}_{\lambda_1{\otimes}{\cdots} {\otimes} \lambda_k}(G) =\sum\limits_{j=1}^k \left(  I^{2p}_{\lambda_j}(G_j) \prod\limits_{i\ne j \atop i=1}^k I^{0}_{\lambda_i}(G_i)\right) = \prod\limits_{i=1}^k |O_{\lambda_i}(G_i)|\cdot \sum\limits_{j=1}^k\langle\lambda_j,\lambda_j\rangle^p. \nonumber
\end{eqnarray}
\end{Proposition}

\begin{proof}
For any group $G=G_1 {\times}\ldots{\times} G_k$, the inner product has the following form:
\begin{equation*}
\langle \lambda_1{\otimes}{\cdots} {\otimes} \lambda_k, \mu_1{\otimes}{\cdots} {\otimes} \mu_k \rangle_G = \langle \lambda_1, \mu_1 \rangle_{G_1}+\ldots+ \langle \lambda_k, \mu_k \rangle_{G_k}.
\end{equation*}

It is easy to verify that Formula (\ref{prodsum}) holds.
\end{proof}

\section{Odd-Degree Indices for Orbits}\label{oddsec}

The odd-order index of an irreducible representation serves as a parameter limiting the symmetry of the mathematical model of particle physics and its diverse extensions \cite{Okubo1}. The triangular anomaly numbers have been defined for the Lie group $SU(n)$ by the sum of cubes of the components of the weights corresponding to the $U(1)$ subgroup in the reduction $SU(n) \supset U(1) {\times} SU(n-1)$ \cite{PS}. 

In general, the crucial part of obtaining the anomaly number lies in determining the vector $v$ passing through the origin of the weight space. For any Coxeter group $H_n$, $n\in\{2,3,4\}$, the orbits of its lower subgroup span $\R^{n-1}$ orthogonal to $v$. Projecting the orbit points onto $v$ and calculating the sum of the distances between the obtained projections, we can determine whether this sum yields zero or not. Generally, the highest weight of the unitary group $U(1)$ sets the direction of $v$. However, other suitable vectors are not excluded, and they are utilized as long as the resulting sum is not equal to zero.

\begin{figure}[H]
\centering
\parbox{.6\linewidth}{\setlength{\unitlength}{1.8pt}
\def\kr{\circle{4}}
\thicklines
\begin{picture}(100,10)
\put(6,10){\makebox(0,0){${H_2}$}}
\put(19,10){\kr}
\put(19,5){\makebox(0,0){$\a_1$}}
\put(29,10){\kr}
\put(29,5){\makebox(0,0){$\a_2$}}
\put(21,10){\line(1,0){6}}
\put(23,11){\small$5$}

\put(49,10){\makebox(0,0){${H_3}$}}
\put(59,10){\kr}
\put(59,5){\makebox(0,0){$\a_1$}}
\put(69,10){\kr}
\put(69,5){\makebox(0,0){$\a_2$}}
\put(79,10){\kr}
\put(79,5){\makebox(0,0){$\a_3$}}
\put(61,10){\line(1,0){6}}
\put(71,10){\line(1,0){6}}
\put(73,11){\small$5$}

\put(99,10){\makebox(0,0){${H_4}$}}
\put(109,10){\kr}
\put(109,5){\makebox(0,0){$\a_1$}}
\put(119,10){\kr}
\put(119,5){\makebox(0,0){$\a_2$}}
\put(129,10){\kr}
\put(129,5){\makebox(0,0){$\a_3$}}
\put(139,10){\kr}
\put(139,5){\makebox(0,0){$\a_4$}}
\put(111,10){\line(1,0){6}}
\put(121,10){\line(1,0){6}}
\put(131,10){\line(1,0){6}}
\put(133,11){\small$5$}
\end{picture}
}
\caption{The Coxeter--Dynkin diagrams of the non-crystallographic groups $H_n$, $n\in\{2,3,4\}$. The~nodes correspond to the simple roots $\alpha_k$, $k\in\{1,\dots,n\}$.}\label{diags}
\end{figure}
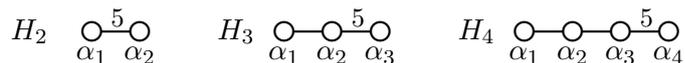

The non-zero anomaly numbers exist only for those groups that have a corresponding symmetric Coxeter--Dynkin diagram. From such diagrams for the non-crystallographic groups (Figure~\ref{diags}), it~follows that the anomaly number of the Coxeter group $H_2$ is not equal to zero. The~non-crystallographic groups $H_3$ and $H_4$ are anomaly-free groups, as their Coxeter--Dynkin diagrams are non-symmetric.

The nodes of the Coxeter--Dynkin diagrams of $H_n$, $n\in\{2,3,4\}$ can be also labeled by the reflections $r_k$ across the hyperplanes orthogonal to $\alpha_k$, $k\in\{1,\dots,n\}$. For a non-crystallographic group $H_n$ and any point $x\in\R^n$, the reflection formula is provided by the scalar product \eqref{inprod} as
\begin{equation}\label{refl}
r_k x = x - \langle x, \alpha_k \rangle, \quad k\in\{1,\dots,n\}.
\end{equation}

\begin{figure}[H]
\centering
\includegraphics[scale=0.27]{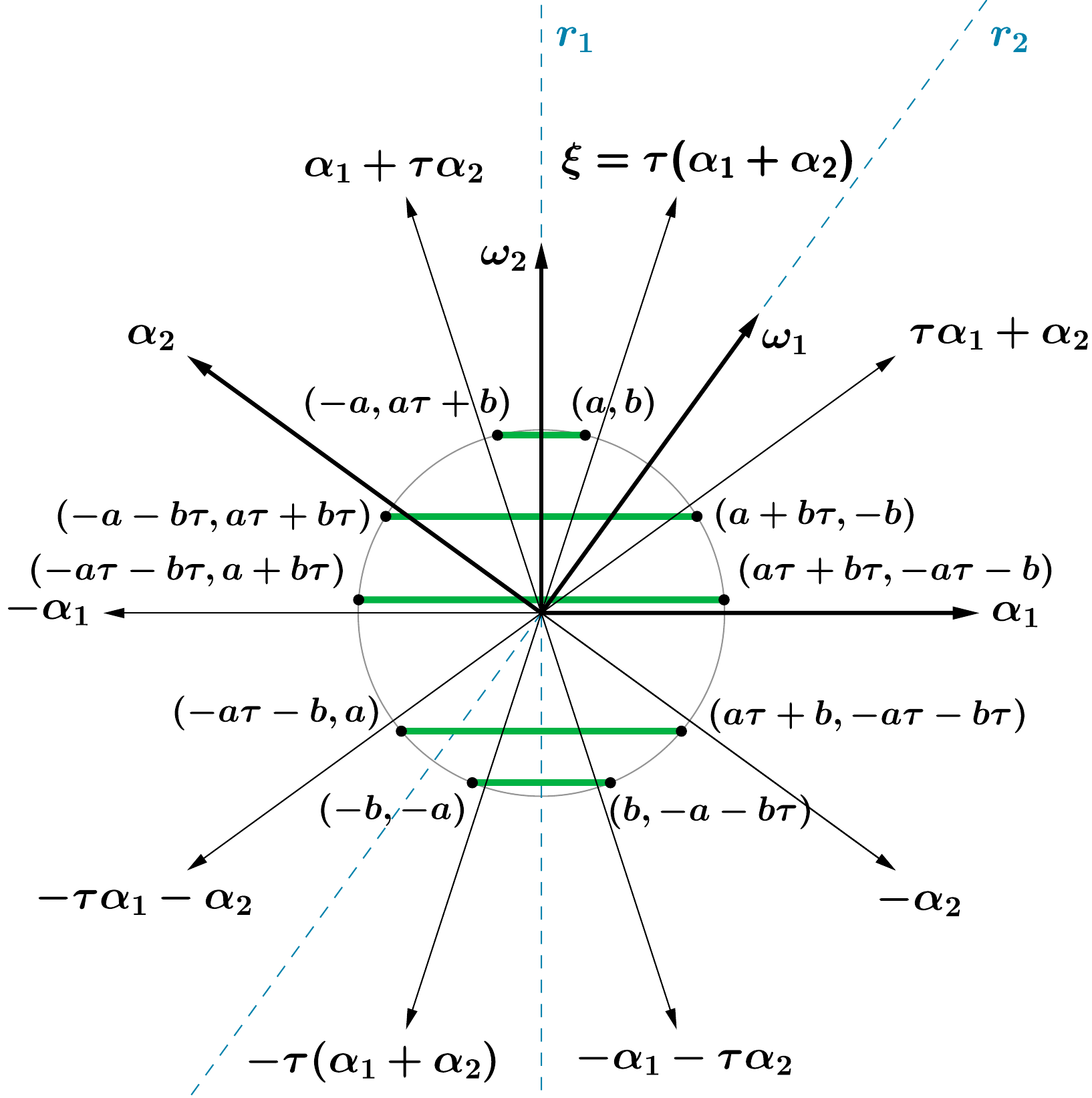}
\caption{The root system of the Coxeter group $H_2$. The dashed lines $r_1$ and $r_2$ correspond to the reflecting hyperplanes orthogonal to the simple roots $\alpha_1$ and $\alpha_2$, respectively. The root $\xi$ denotes the highest root of $H_2$. The coordinates of the points of an orbit with a dominant point $\lambda=(a,b)$ of $H_2$ are listed. The orbits of the reflection group $A_1$ are depicted by green segments.}\label{fig2}
\end{figure}

\begin{Definition}\label{odd}
Let $G$ be a Coxeter group $H_n$, $n\in\{2,3,4\}$ of non-crystallographic type. The number defined by 
\begin{equation}\label{ann}
A_\lambda^{2p-1}(G)=\sum_{\mu\in O_\lambda(G)} \langle \mu, v \rangle^{2p-1}, \quad p\in\mathbb{N},
\end{equation}
where $v$ is a vector orthogonal to the simple roots $\alpha_1,\ldots, \alpha_{k-1}$, is called the {anomaly number} or {$(2p-1)$-th-order index} of an orbit $O_\lambda(H_n)$.
\end{Definition}

\begin{Example}
Consider an orbit of the non-crystallographic group $H_{2}$ with a dominant point $\lambda=(a,b)$ shown in Figure~\ref{fig2}. In this case, the weight $\omega_2$ can be chosen as the vector $v$, since it is orthogonal to the simple root $\alpha_1$. Hence, using Formula (\ref{ann}), the calculations of the anomaly numbers yield:
\begin{align*}
A_{(a,b)}^{2p {-}1}(H_2)=&\sum_{\mu\in O_{(a,b)}(H_2)} \langle \mu, (0,1) \rangle^{2p-1}=
2\left(\frac{1}{\tau{-}3}\right)^{2p{-}1}\left\{ (a(\tau{-}1) {-}b(\tau{-}1))^{2p {-}1}\right.\\
& {-}(a \tau {+}2b)^{2p {-}1} {-}(a \tau {+}b(\tau{-}1))^{2p {-}1}
\left. {+}(2a{+}b\tau)^{2p {-}1} {+}(a(\tau{-}1) {+}b\tau)^{2p {-}1}\right\}.
\end{align*}
\end{Example}

\begin{Remark}\label{anrem}\
We can generalize Definition~\ref{odd} by taking into consideration the following statements:
\begin{itemize}
\item The anomaly numbers $A^1_{(a,b)}(H_2) =A^3_{(a,b)}(H_2) = 0, \textrm{ for any } a,b\in \mathbb{R}.$ 
\item The odd-order indices $A^{2p-1}_{(a,b)}(H_2)\ne 0$, for $a\ne b$ and $p>2$.
\item  For the Coxeter groups $H_3$ and $H_4$, as any orbit contains the elements with positive and negative signs, the anomaly numbers obtained for any orbit are equal to zero:
\begin{equation*}
A^{2p-1}_{\lambda}(H_n)=\sum_{\mu\in O_{\lambda}(H_n)} \langle \mu, v \rangle^{2p-1}=0, \quad n=3,4.
\end{equation*}
\end{itemize}
\end{Remark}

\begin{Definition}\label{ano}
Let $G$ be a non-crystallographic finite reflection group $H_2$. The number defined by
\begin{equation*}
A_\lambda^{p}(H_2)=\sum_{\mu\in O_\lambda(H_2)} \langle \mu, \omega_2 \rangle^{p}, \quad p\in\mathbb{N}\cup\{0\}
\end{equation*}
is called the {$p$-th-order anomaly number} of an orbit $O_\lambda(H_2)$.
\end{Definition}

From Definition~\ref{ano}, one can immediately notice the following relation:
\begin{equation*}
A^0_{\lambda}(H_2)=|O_\lambda(H_2)|.
\end{equation*}

The general formulas for $p$-th-order anomaly number of any orbit of the $H_2$ group are given by:
\begin{align}\label{anoH2}
(3-\tau)^pA^p_{(a,b)}(H_2)&= 2[((a{-}b)(1{-}\tau))^p{+}(a\tau{+}2b)^p{+}({-}2a{-}b\tau)^p \\ \nonumber
&{+}(a(1{-}\tau){-}b\tau)^p{+}(a\tau{+}b(\tau{-}1))^p], \quad a,b\ne 0, \\\nonumber
(3-\tau)^pA^p_{(a,0)}(H_2)&= 2[(a(1{-}\tau))^p{+}(a\tau)^p]{+}({-}2a)^p, \quad a\ne0, \\ \nonumber
(3-\tau)^pA^p_{(0,b)}(H_2)&= 2[(b(\tau{-}1))^p{+}({-}b\tau)^p]{+}(2b)^p, \quad b\ne0.  \nonumber
\end{align}

Comparing the formulas for the lower even-order indices (Proposition~\ref{indH2}) and for $p$-th-order anomaly numbers \eqref{anoH2}, one can observe the following equalities:
\begin{align*}
A^0_\lambda(H_2)&=I^0_\lambda(H_2),\\
A^2_\lambda(H_2)&=\frac{1}{3-\tau}I^2_\lambda(H_2),\\
A^4_\lambda(H_2)&=\frac{3}{2(3-\tau)^2}I^4_\lambda(H_2),\\
A^6_\lambda(H_2)&=\frac{5}{2(3-\tau)^3}I^6_\lambda(H_2),\\
A^8_\lambda(H_2)&=\frac{35}{8(3-\tau)^4}I^8_\lambda(H_2).
\end{align*}

Similarly to the even-order indices, the formulas for the direct sum and tensor product can be derived for the anomaly numbers of orbits for the Coxeter group $H_2$.
\begin{Proposition}
The formula for the $p$-th-order anomaly number of the direct sum of $k$ orbits $\lambda_i\in O_{\lambda_i}(H_2)$, $i\in\{1,2,\ldots,k\}$ is given by:
\begin{equation*}
A^{p}_{{\lambda_1\oplus \ldots\oplus \lambda_k}}(H_2)=\sum_{i=1}^k A^{p}_{\lambda_i}(H_2).
\end{equation*}

The formulas for the $p$-th-order anomaly numbers of the tensor product of two and three orbits $\lambda_i\in O_{\lambda_i}(H_2)$, $i\in\{1,2,3\}$ are given by:
\begin{align*}
A^{p}_{{\lambda_1\otimes  \lambda_2}}(H_2)&=\sum_{i=0}^2 \left(p \atop i \right) A^{i}_{\lambda_1}(H_2)A^{p-i}_{\lambda_2}(H_2),\\
A^{p}_{\lambda_1{\otimes}\lambda_2 {\otimes} \lambda_3}(H_2) 
&=\sum_{i=0}^3 \left(p \atop i \right) A^{i}_{\lambda_1}(H_2) \sum_{j=0}^{3-i} \left(p{-}i \atop j \right) A^{j}_{\lambda_2}(H_2)A^{p-i-j}_{\lambda_3}(H_2).
\end{align*}
\end{Proposition}

\begin{Remark}
In general, the $p$-th-order anomaly numbers of $k$-th tensor product of orbits of the $H_2$ group are defined recursively as follows:
\begin{equation*}
A^{p}_{\lambda_1{\otimes}{\cdots} {\otimes} \lambda_k}(H_2) = A^{p}_{\lambda_1{\otimes}( \lambda_2\otimes{\cdots} {\otimes} \lambda_k)}(H_2) ,\quad  k\in \mathbb{N}^{\geq2}.
\end{equation*}
\end{Remark}

\section{Embedding Index}\label{embsec} 

In order to determine the embedding index of an irreducible representation, the branching rule should be defined for a given Lie algebra and its subalgebra \cite{Okubo2}. Such rules have been calculated for numerous irreducible representations of simple Lie algebras \cite{McKP}. Applying the branching rule to any orbit of a non-crystallographic reflection group, we can reduce any chosen orbit to a sum of several orbits. Such a decomposition corresponds to the subgroups of a chosen Coxeter group. Dividing the size of an orbit of any Coxeter group by the size of its reduced orbit provides a specific ratio called the embedding index. 

The index considered in this section depends only on the rank $r$ of a finite reflection group. Whenever any branching rule is established, it takes the same value for all orbits. Given that the embedding index can be obtained for any orbit of a crystallographic reflection group, we demonstrate that this property holds for the non-crystallographic groups $H_n$, $n\in\{2,3,4\}$ as well.

\begin{Definition}
Let $G$ be a reflection group of order $n$, and $G_1 {\times}\ldots{\times} G_k$, $k \le n$, be a maximal subgroup of $G$. The second-order index of the embedding $G \hookleftarrow  G_1 {\times}\ldots{\times} G_k$ is given by the formula:
\begin{equation}\label{embindex}
\gamma=\frac{I^{2}(G)}{I^{2}(G_1{\times}\ldots{\times} G_k)}.
\end{equation}
\end{Definition}

\begin{Remark}
The formula for the embedding index is generalized for any parameter $k$. However, in this paper, we~focus only on the non-crystallographic cases with $k\in\{2,3,4\}$.
\end{Remark}

Using the Formula (\ref{embindex}), we calculate the embedding indices $\gamma$ for any Coxeter group of non-crystallographic type and its maximal subgroup (Table~\ref{nonindex}).

\begin{table}[ht]
\caption{The embedding index $\gamma$ provided for the non-crystallographic groups $H_n$, $n\in\{2,3,4\}$ and their maximal subgroups $G'$.} \label{nonindex}
\centering
\begin{tabular}{c|c|c}
\boldmath{$G$} & \boldmath{$G'$}& \boldmath{$\gamma$}\\ \hline
$H_{2}$ & $ A_{1}$ & $2$ \\
$H_3$  & $A_{1} {\times} A_{1} {\times} A_{1}$ & $1$ \\
$H_3$  & $A_{2}$ & $3/2$ \\
$H_3$ & $ H_{2}$ & $3/2$ \\
\end{tabular}\qquad
\begin{tabular}{c|c|c}
\boldmath{$G$} & \boldmath{$G'$}& \boldmath{$\gamma$} \\ \hline
$H_4$ & $ A_{2} {\times} A_{2}$ & $1$ \\
$H_4$ & $ H_{2} {\times} H_{2}$ & $1$ \\
$H_4$ & $ A_{1} {\times} A_{1} {\times} A_{1}{\times} A_{1}$ & $1$ \\ 
$H_4$ & $ H_{3} {\times} A_{1}$ & $1$ \\
$H_4$ & $ A_{4}$ & $1$ \\
$H_4$ & $D_{4}$ & $1$ \\
\end{tabular}
\end{table}

\begin{Theorem}
For any Coxeter group $G$ of non-crystallographic type, the embedding index $\gamma$  is a fraction of the ranks, i.e., those of a group $G$ and its maximal subgroup $G'$, namely:
\begin{equation*}
\gamma=\frac{\textrm{rank } G}{\textrm{rank } G'}.
\end{equation*}
\end{Theorem}

\begin{proof}
Let us consider the two cases: $(i)$ $\textrm{rank } G = \textrm{rank } G'$, and $(ii)$ $\textrm{rank } G > \textrm{rank} G'$.

$(i)$ The elements of any orbit $O_\lambda(G)$ of a group $G$ are found on the surface of a sphere with a finite radius. Applying the branching rule method to $\lambda$, we obtain several orbits of the subgroup $G'$ of a group $G$. Since $\textrm{rank } G' = \textrm{rank } G$, all elements of orbits of $G'$ are found on the surface of a sphere of the same radius. 
Since the second-order index is given by the sum of squared distances between the orbit points and the origin, we have that $I^2(G)=I^2(G')$. In such a case, the index $\gamma$ is equal to $1$. 

$(ii)$ First, let us recall that for any orbit $O_\lambda(H_3)$, the orbit points have the coordinates $(x,y,z)$ in the $\omega-$basis. In this case, some particular values occur an equal number of times for $x,y$ and $z$. This~property arises due to the impact of the tetrahedral symmetry of the non-crystallographic $H_3$ group on the orbit points.

For instance, the orbits of any maximal subgroup $G'$ of the Coxeter group $H_3$ are selected in the following way: 
\begin{itemize}
\item consider the points of an orbit $O_{\lambda}(H_3)$; 
\item remove the first coordinate of each point in the case of $H_2$, and the third one for the crystallographic group $A_2$; 
\item among all the points in $\R^2$ select those with non-negative coordinates; such points provide the orbits of $G'$ in $\R^2$. 
\end{itemize}

Considering the values appearing at each coordinate, the index $I^2$ of the subgroup $G'$ of $H_3$ is equal to $\frac23 I^2(H_3)$. Therefore, the embedding index $\gamma=\frac32.$ A similar explanation can be provided for the $H_2$ group.
\end{proof}

\section{Lower Orbits of $H_2$ and $H_3$}\label{detsec}

For simple Lie algebras, using the highest weight of an irreducible representation, we~can determine its dominant weight multiplicities by subtracting the simple roots \cite{Brem}. Hence, the~computational problem comprises the following components: 
\begin{itemize}
\item determination of the highest weight; 
\item subtraction of simple roots from the highest weight;
\item an algorithm that describes the subtraction path. 
\end{itemize}

For crystallographic cases, the appearance of dominant weight multiplicities arrises from the non-commutativity of the certain elements of a Lie algebra \cite{MP82, Brem1}. Since, in the case of finite reflection groups, all reflections commute, a similar procedure can be developed and properly applied to individual orbits of the considered non-crystallographic groups. The multiple occurrences of equal dominant weights within one system necessarily involve the same number of dominant points of corresponding lower orbits. 

In this section, we only examine the groups $H_n$, $n\in\{2,3\}$; their simple roots $\alpha_i$, $i\in\{1,\dots,n\}$ are provided by the Cartan matrices (Table~\ref{Cartan}). In order to identify all lower orbits of $H_2$ and $H_3$, the~algorithm contains the following steps:

\begin{description}
\item[$(i)$] determine a dominant point $\lambda=(l_1,\dots,l_i)$, $l_i=a_i+b_i\tau \in \mathbb{Z}[\tau]^{>0}$, $i\in\{1,\dots,n\}$;
\vspace{1.5mm}
\item[$(ii)$] establish a correspondence between the coordinates of a dominant point $\lambda$ and the index\newline $i\in\{1,\dots,n\}$ of a simple root $\alpha_i$: $i \rightarrow l_i$;
\vspace{1.5mm}
\item[$(iii)$] if at least one of $l_i>0$, $i\in\{1,\dots,n\}$, then proceed the following subtraction: 
\begin{itemize}
\item if $b_i=0$, then $\mu_i=\lambda-j\cdot\alpha_i$, $j\in\{1,\ldots,a_i\};$
\item if $b_i\ge1$:
\begin{itemize}
\item and $a_i=0$, then $\mu_i=\lambda-k\tau\cdot\alpha_i$, $k\in\{1,\ldots,b_i\};$
\item and $a_i\ge1$, then $\mu_i=\lambda-k\frac{l_i}{gcd(a_i,b_i)}\cdot\alpha_i$, $ k\in\{1,2,\ldots,gcd(a_i,b_i)\};$
\end{itemize}
\end{itemize}
\item[$(iv)$] replace a point $\lambda$ in $(i)$ with $\mu_i$;
\item[$(v)$] repeat the steps $(ii)$--$(iv)$ until at least one of the coordinates $\mu_i$ is greater than zero.
\end{description}

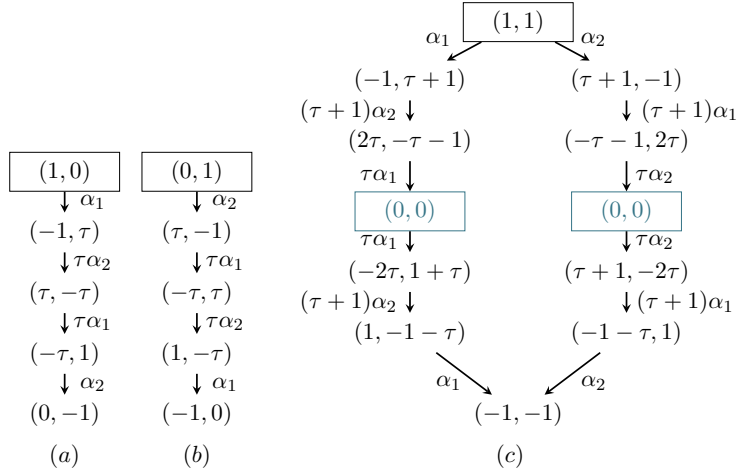
\begin{figure}[H]
\centering
\resizebox{.63\textwidth}{!}{
\begin{tabular}{ccc}
\begin{tikzpicture}
\node[block](start) {$(1,0)$};
\node[below of=start] (c1) {$(-1,\tau)$};
\node[below of=c1] (c2) {$(\tau,-\tau)$};
\node[below of=c2] (c3) {$(-\tau,1)$};
\node[below of=c3] (c4) {$(0,-1)$};
\path [line] (start) -- node[yshift=0em, xshift=0.8em] {$\hspace{3mm}\a_1$} (c1);
\path [line] (c1) -- node[yshift=0em, xshift=0.8em] {$\hspace{3mm}\tau\a_2$} (c2);
\path [line] (c2) -- node[yshift=0em, xshift=0.8em] {$\hspace{3mm}\tau\a_1$} (c3);
\path [line] (c3) -- node[yshift=0em, xshift=0.8em] {$\hspace{3mm}\a_2$} (c4);
\end{tikzpicture} &
\begin{tikzpicture}
\node[block](start) {$(0,1)$};
\node[below of=start] (c1) {$(\tau,-1)$};
\node[below of=c1] (c2) {$(-\tau,\tau)$};
\node[below of=c2] (c3) {$(1,-\tau)$};
\node[below of=c3] (c4) {$(-1,0)$};
\path [line] (start) -- node[yshift=0em, xshift=0.8em] {$\hspace{3mm}\a_2$} (c1);
\path [line] (c1) -- node[yshift=0em, xshift=0.8em] {$\hspace{3mm}\tau\a_1$} (c2);
\path [line] (c2) -- node[yshift=0em, xshift=0.8em] {$\hspace{3mm}\tau\a_2$} (c3);
\path [line] (c3) -- node[yshift=0em, xshift=0.8em] {$\hspace{3mm}\a_1$} (c4);
\end{tikzpicture} &
\begin{tikzpicture}
\node[block](start) {$(1,1)$};
\node[left of=start, xshift=-2em,yshift=-2.5em] (c1) {$(-1,\tau+1)$};
\node[right of=start, xshift=2em,yshift=-2.5em] (c11) {$(\tau+1,-1)$};
\node[below of=c1, yshift=0em] (c2) {$(2\tau,-\tau-1)$};
\node[block,color=mycolor,below of=c2, yshift=0em] (c21) {$(0,0)$};
\node[below of=c21, yshift=0em] (c22) {$(-2\tau,1+\tau)$};
\node[below of=c22, yshift=0em] (c23) {$(1,-1-\tau)$};
\node[below of=c11, yshift=0em] (c3) {$(-\tau-1,2\tau)$};
\node[block,color=mycolor,below of=c3, yshift=0em] (c31) {$(0,0)$};
\node[below of=c31, yshift=0em] (c32) {$(\tau+1,-2\tau)$};
\node[below of=c32, yshift=0em] (c33) {$(-1-\tau,1)$};
\node[below of=start, yshift=-14.2em] (c4) {$(-1,-1)$};
\path [line] (start) -- node[yshift=0.5em, xshift=-0.7em] {$\hspace{-3mm}\a_1$} (c1) ;
\path [line] (c1) -- node[yshift=0em, xshift=-3em] {$\hspace{3mm}(\tau+1)\a_2$} (c2);
\path [line] (c2) -- node[yshift=0em, xshift=-1.7em] {$\hspace{3mm}\tau\a_1$} (c21);
\path [line] (c21) -- node[yshift=0em, xshift=-1.7em] {$\hspace{3mm}\tau\a_1$} (c22);
\path [line] (c22) -- node[yshift=0em, xshift=-3em] {$\hspace{3mm}(\tau+1)\a_2$} (c23);
\path [line] (start) -- node[yshift=0.5em, xshift=0.5em] {$\hspace{3mm}\a_2$} (c11);
\path [line] (c11) -- node[yshift=0em, xshift=2.3em] {$\hspace{3mm}(\tau+1)\a_1$} (c3);
\path [line] (c3) -- node[yshift=0em, xshift=0.8em] {$\hspace{3mm}\tau\a_2$} (c31);
\path [line] (c31) -- node[yshift=0em, xshift=0.8em] {$\hspace{3mm}\tau\a_2$} (c32);
\path [line] (c32) -- node[yshift=0em, xshift=2.1em] {$\hspace{3mm}(\tau+1)\a_1$} (c33);
\path [line] (c33) -- node[yshift=-0.4em, xshift=-1.3em] {$\hspace{17mm}\a_2$} (c4);
\path [line] (c23) -- node[yshift=-0.4em, xshift=1.3em] {$\hspace{-15mm}\a_1$} (c4);
\end{tikzpicture} \\
$(a)$ & $(b)$ & $(c)$ \\
\end{tabular}
}
\caption{The tree-diagram for the orbits of $H_2$. (\textbf{a}) $O_{(1,0)}(H_2)$; (\textbf{b}) $O_{(0,1)}(H_2)$; (\textbf{c}) $O_{(1,1)}(H_2)$. The~dominant points are displayed in boxes. The points that do not belong to $O_{(1,1)}(H_2)$ are depicted by blue colour.}\label{H2subpath}
\end{figure}

\begin{figure}[H]
\centering
\resizebox{.46\textwidth}{!}{
\begin{tikzpicture}
nodes
\node[block](start) {$(\tau,1)$};
\node[left of=start, xshift=-2em,yshift=-2.5em] (c1) {$(-\tau,\tau+2)$};
\node[right of=start, xshift=2em,yshift=-2.5em] (c11) {$(2\tau,-1)$};
\node[below of=c1, yshift=0em] (c2) {$(2\tau+1,-\tau-2)$};
\node[below of=c2, yshift=0em] (c21) {$(-2\tau-1,2\tau)$};
\node[below of=c21,color=mycolor, yshift=0em] (c22) {$(-\tau,0)$};
\node[below of=c22, yshift=0em] (c23) {$(1,-2\tau)$};
\node[block,color=mycolor,below of=c11, yshift=0em] (c3) {$(0,\tau)$};
\node[below of=c3, yshift=0em] (c31) {$(-2\tau,2\tau+1)$};
\node[below of=c31, yshift=0em] (c32) {$(\tau+2,-2\tau-1)$};
\node[below of=c32, yshift=0em] (c33) {$(-\tau-2,\tau)$};
\node[below of=start, yshift=-13.5em] (c4) {$(-1,-\tau)$};
\path [line] (start) -- node[yshift=0.5em, xshift=-0.7em] {$\hspace{-3mm}\tau\a_1$} (c1) ;
\path [line] (c1) -- node[yshift=0em, xshift=-3em] {$\hspace{3mm}(\tau+2)\a_2$} (c2);
\path [line] (c2) -- node[yshift=0em, xshift=-3.1em] {$\hspace{3mm}(2\tau+1)\a_1$} (c21);
\path [line] (c21) -- node[yshift=0em, xshift=-1.7em] {$\hspace{3mm}\tau\a_2$} (c22);
\path [line] (c22) -- node[yshift=0em, xshift=-1.7em] {$\hspace{3mm}\tau\a_2$} (c23);
\path [line] (start) -- node[yshift=0.5em, xshift=0.5em] {$\hspace{3mm}\a_2$} (c11);
\path [line] (c11) -- node[yshift=0em, xshift=0.8em] {$\hspace{3mm}\tau\a_1$} (c3);
\path [line] (c3) -- node[yshift=0em, xshift=0.8em] {$\hspace{3mm}\tau\a_1$} (c31);
\path [line] (c31) -- node[yshift=0em, xshift=2.3em] {$\hspace{3mm}(2\tau+1)\a_2$} (c32);
\path [line] (c32) -- node[yshift=0em, xshift=2.1em] {$\hspace{3mm}(\tau+2)\a_1$} (c33);
\path [line] (c33) -- node[yshift=-0.4em, xshift=-1.3em] {$\hspace{17mm}\tau\a_2$} (c4);
\path [line] (c23) -- node[yshift=-0.4em, xshift=1.3em] {$\hspace{-15mm}\a_1$} (c4);
\end{tikzpicture}
}
\caption{The tree-diagram for the orbit $O_{(\tau,1)}(H_2)$. The dominant points are displayed in boxes. The~points that do not belong to $O_{(\tau,1)}(H_2)$ are depicted by blue colour.}\label{multpH2}
\end{figure}
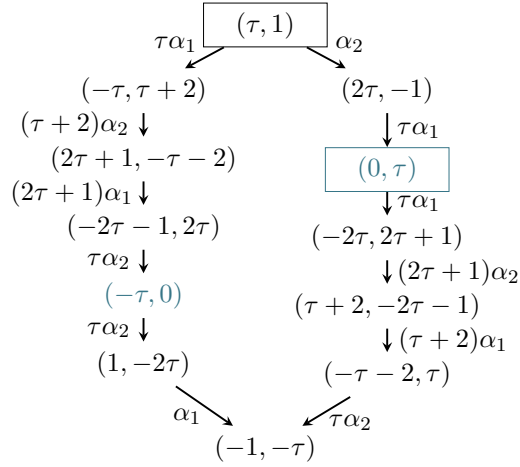

\begin{figure}[H]
\centering
\begin{tabular}{ccc}
\resizebox{.3\textwidth}{!}{
\begin{tikzpicture}
\node [block](start) {$(1,0,0)$};
\node[below of=start, yshift=0em]  (c1) {$(-1,1,0)$};
\node[below of=c1, yshift=0em] (c2) {$(0,-1,\tau)$};
\node[below of=c2, yshift=0em] (c3) {$(0,\tau,-\tau)$};
\node[below of=c3, yshift=0em] (c4) {$(\tau,-\tau,1)$};
\node[left of=c4, xshift=-2em,yshift=-2.5em] (c5) {$(-\tau,0,1)$};
\node[right of=c4, xshift=2em,yshift=-2.5em] (c51) {$(\tau,0,-1)$};
\node[below of=c4, yshift=-2.3em] (c6) {$(-\tau,\tau,-1)$};
\node[below of=c4, yshift=-5em] (c8) {$(0,-\tau,\tau)$};
\node[below of=c8, yshift=0em] (c9) {$(0,1,-\tau)$};
\node[below of=c9, yshift=0em] (c10) {$(1,-1,0)$};
\node[below of=c10, yshift=0em] (c11) {$(-1,0,0)$};
\path [line] (start) -- node[yshift=0em, xshift=0.8em] {$\hspace{3mm}\a_1$} (c1);
\path [line] (c1) -- node[yshift=0em, xshift=0.8em] {$\hspace{3mm}\a_2$} (c2);
\path [line] (c2) -- node[yshift=0em, xshift=0.8em] {$\hspace{3mm}\tau\a_3$} (c3);
\path [line] (c3) -- node[yshift=0em, xshift=0.8em] {$\hspace{3mm}\tau\a_2$} (c4);
\path [line] (c4) -- node[yshift=0.5em, xshift=-0.3em] {$\hspace{-3mm}\tau\a_1$} (c5) ;
\path [line] (c5) -- node[yshift=-0.3em, xshift=-0.8em] {$\hspace{0mm}\a_3$} (c6);
\path [line] (c4) -- node[yshift=0.5em, xshift=0.3em] {$\hspace{3mm}\a_3$} (c51);
\path [line,color=mycolor] (c51) -- node[yshift=-0.3em, xshift=0.8em] {$\hspace{3mm}\tau\a_1$} (c6);
\path [line] (c6) -- node[yshift=0em, xshift=0.8em] {$\hspace{3mm}\tau\a_2$} (c8);
\path [line] (c8) -- node[yshift=0em, xshift=0.8em] {$\hspace{3mm}\tau\a_3$} (c9);
\path [line] (c9) -- node[yshift=0em, xshift=0.8em] {$\hspace{3mm}\a_2$} (c10);
\path [line] (c10) -- node[yshift=0em, xshift=0.8em] {$\hspace{3mm}\a_1$} (c11);
\end{tikzpicture}} & \hspace{2mm} &
\resizebox{.65\textwidth}{!}{
\begin{tikzpicture}
\node [block](start) {$(0,0,1)$};
\node[below of=start]  (c2) {$(0,\tau,-1)$};
\node[below of=c2] (c3) {$(\tau,-\tau,\tau)$};
\node[left of=c3, xshift=-2em,yshift=-3em] (c4) {$(-\tau,0,\tau)$};
\node[right of=c3,  xshift=2em,yshift=-3em] (c41) {$(\tau,1,-\tau)$};
\node[right of=c41,  xshift=2em,yshift=-2.5em] (c42) {$(\tau+1,-1,0)$};
\node[below of=c42] (c43) {$(-\tau-1,\tau,0)$};
\node[below of=c4] (c5) {$(-\tau,\tau+1,-\tau)$};
\node[below of=c5] (c6) {$(1,-\tau-1,\tau+1)$}; 
\node[left of=c6, xshift=-2em,yshift=-2.8em] (c7) {$(1,\tau,-\tau-1)$};
\node[right of=c6, xshift=2em,yshift=-2.8em] (c71) {$(-1,-\tau,\tau+1)$};
\node[left of=c7, xshift=-2em,yshift=-2.8em] (c72) {$(\tau+1,-\tau,0)$};
\node[right of=c7, xshift=2em,yshift=-2.8em] (c73) {$(-1,\tau+1,-\tau-1)$};
\node[below of=c72] (c74) {$(-\tau-1,1,0)$};
\node[below of=c73] (c75) {$(\tau,-\tau-1,\tau)$};
\node[below of=c74, xshift=5em] (c76) {$(-\tau,-1,\tau)$};
\node[below of=c74, xshift=14em] (c77) {$(\tau,0,-\tau)$}; 
\node[below of=c75, yshift=-2.8em] (c78) {$(-\tau,\tau,-\tau)$};
\node[below of=c78] (c79) {$(0,-\tau,1)$};
\node[below of=c79] (c80) {$(0,0,-1)$};
\path [line] (start) -- node[yshift=0em, xshift=0.8em] {$\hspace{3mm}\a_3$} (c1);
\path [line] (c1) -- node[yshift=0em, xshift=0.8em] {$\hspace{3mm}\tau\a_2$} (c3);
\path [line] (c3) -- node[yshift=0.5em, xshift=-0.3em] {$\hspace{-3mm}\tau\a_1$} (c4);
\path [line] (c3) -- node[yshift=0.4em, xshift=1.3em] {$\hspace{-3mm}\tau\a_3$} (c41);
\path [line] (c4) -- node[yshift=0em, xshift=0.8em] {$\hspace{3mm}\tau\a_3$} (c5);
\path [line,color=mycolor] (c41) -- node[yshift=0.6em, xshift=-0.5em] {$\hspace{3mm}\tau\a_1$} (c5);
\path [line] (c41) -- node[yshift=0.5em, xshift=0em] {$\hspace{3mm}\a_2$} (c42);
\path [line] (c42) -- node[yshift=0em, xshift=2em] {$\hspace{3mm}(\tau+1)\a_1$} (c43);
\path [line,color=mycolor] (c43) -- node[yshift=0.6em, xshift=-0.6em] {$\hspace{3mm}\tau\a_2$} (c71);
\path [line] (c5) -- node[yshift=0em, xshift=2em] {$\hspace{3mm}(\tau+1)\a_2$} (c6);
\path [line] (c6) -- node[yshift=0.3em, xshift=-0.3em] {$\hspace{-20mm}(\tau+1)\a_3$} (c7);
\path [line] (c6) -- node[yshift=0.3em, xshift=2em] {$\hspace{-7mm}\a_1$} (c71);
\path [line] (c7) -- node[yshift=0.3em, xshift=1em] {$\hspace{-20mm}\tau\a_2$} (c72);
\path [line] (c7) -- node[yshift=0.3em, xshift=3.7em] {$\hspace{-20mm}\a_1$} (c73);
\path [line,color=mycolor] (c71) -- node[yshift=0em, xshift=5.5em] {$\hspace{-20mm}(\tau+1)\a_3$} (c73);
\path [line] (c72) -- node[yshift=0em, xshift=5em] {$\hspace{-20mm}(\tau+1)\a_1$} (c74);
\path [line] (c73) -- node[yshift=0em, xshift=5em] {$\hspace{-20mm}(\tau+1)\a_2$} (c75);
\path [line] (c74) -- node[yshift=-0.3em, xshift=1.7em] {$\hspace{-20mm}\a_2$} (c76);
\path [line,color=mycolor] (c75) -- node[yshift=-0.3em, xshift=4em] {$\hspace{-20mm}\tau\a_1$} (c76);
\path [line] (c75) -- node[yshift=0.3em, xshift=3.7em] {$\hspace{-20mm}\tau\a_3$} (c77);
\path [line] (c76) -- node[yshift=0em, xshift=1.1em] {$\hspace{-20mm}\tau\a_3$} (c78);
\path [line,color=mycolor] (c77) -- node[yshift=0em, xshift=4.3em] {$\hspace{-20mm}\tau\a_1$} (c78);
\path [line] (c78) -- node[yshift=0em, xshift=3.8em] {$\hspace{-20mm}\tau\a_2$} (c79);
\path [line] (c79) -- node[yshift=0em, xshift=3.8em] {$\hspace{-20mm}\a_3$} (c80);
\end{tikzpicture}
}\\
$(\textbf{a})$ & \hspace{2mm} & $(\textbf{b})$ \\
\end{tabular}
\caption{The tree-diagrams constructed for the orbits of $H_3$. (\textbf{a}) $O_{(1,0,0)}(H_3)$; (\textbf{b}) $O_{(0,0,1)}(H_3)$. The subtraction paths that yield already existing points are marked by blue color.}
\label{multpH3}
\end{figure}
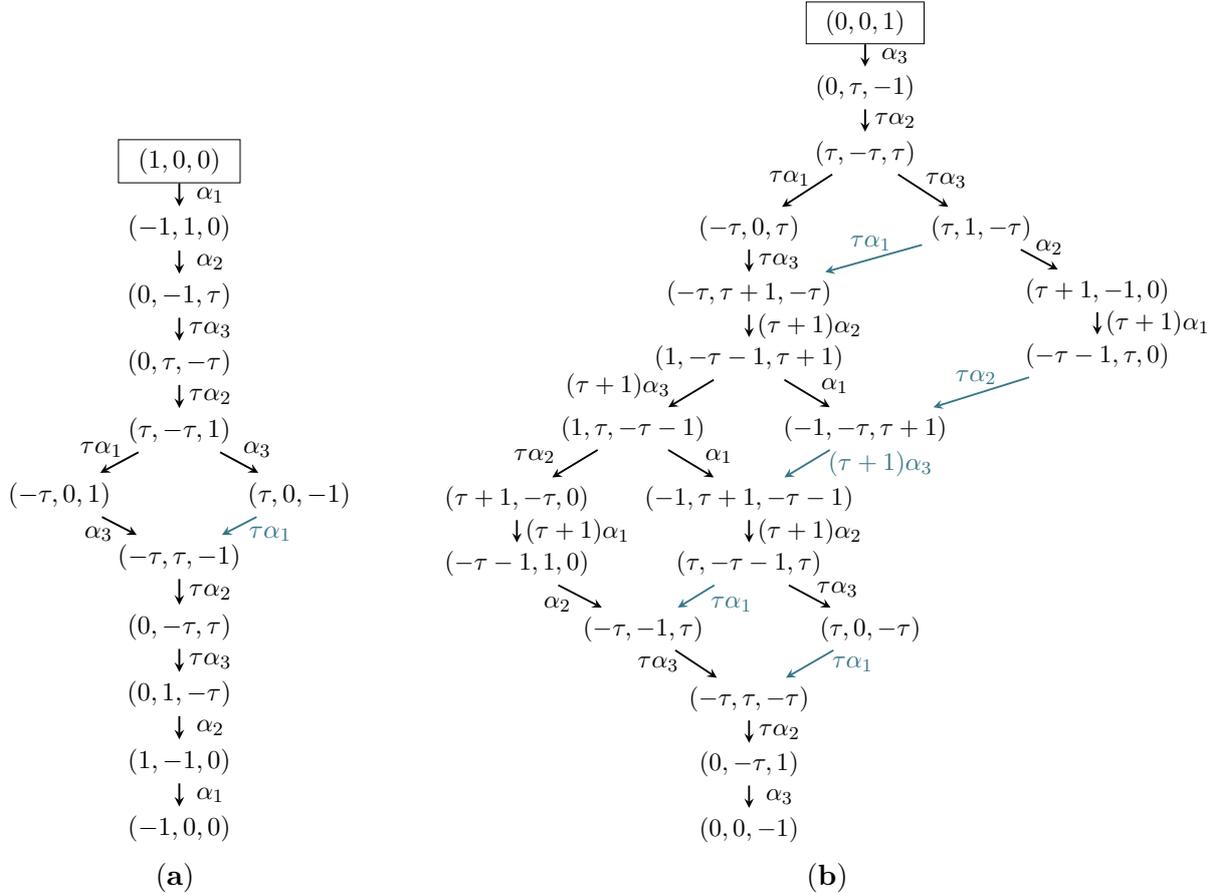

This recursive method provides a tree-diagram for any dominant point $\lambda$ of the $H_2$ and $H_3$ groups (Figures~\ref{H2subpath},~\ref{multpH2} and \ref{multpH3}). Such a method allows one to determine the coordinates of dominant points of lower orbits starting from any chosen $\lambda$. In order to generalize the formulas for the coordinates, it is convenient to consider the dominant points with their coordinates provided by integer coefficients. Furthermore, to obtain such expressions, we only consider the coordinates of dominant points $\lambda$ with equal `dynamic' coefficients. For example, in the case of $H_3$, if $\lambda=(a,b,0)$, it is necessary to consider $a=b$. However, for $\lambda=(a,b,c)$ and $a,b,c>0$, the number of vertices of a corresponding orbit is $|O_{(a,b,c)}(H_3)|=120$, and the generalization of the coordinates of lower orbits becomes less apparent. Therefore, this case is omitted in this paper.

Let us consider a dominant point $\lambda=(a,0)$ of $H_2$. Hence, we can generalize the coordinates of obtained seed points of lower orbits as follows:
\begin{align}\label{H2mult1}
& (a,0) & \quad a\in\N;\\ \nonumber
& \left(a{-}2k{-}2,(k{+}1)\tau\right), \quad k\in\left\{0,\dots,\left[\frac{a}{2}\right]{-}1\right\} & \quad a\in\N^{\geq2};\\ \nonumber
& \left(\frac{a{-}2k}{2}\tau{-}\frac{a{+}2k}{2},2k\tau\right), \quad k\in\left\{0,\dots,\left[\frac{a}{10}\right]\right\} & \quad a=2n,\:n\in\N;\\ \nonumber
& \left(\frac{a{-}2k{-}1}{2}\tau{-}\frac{a{+}2k{+}1}{2},(2k{+}1)\tau\right), \quad k\in\left\{0,\dots,\left[\frac{a{-}2}{10}\right]\right\} & \quad a=2n{+}1,\:n\in\N^{\geq2}. \nonumber
\end{align}

For a dominant point $\lambda=(a,a)$ of $H_2$, only half of the dominant points of lower orbits are provided as
\begin{align}\label{H2mult2}
& (a,a), (0,0) & \quad a\in\N;\\ \nonumber
& \left(a{-}2k,a{+}k\tau\right), \quad k\in\left\{1,\dots,\left[\frac{a}{2}\right]\right\} & \quad a\in\N^{\geq2};\\ \nonumber
& \left((a{-}2k)\tau{-}\frac{a{+}2k}{2},2k(\tau{+}1)\right), \quad k\in\left\{0,\dots,\left[\frac{a{-}n}{2}\right]\right\} & \quad a=2n,\:n\in\N;\\ \nonumber
& \left((a{-}2k{-}1)\tau{-}\frac{a{+}2k{+}1}{2},(2k{+}1)\tau\right), \quad k\in\left\{0,\dots,\left[\frac{a{-}n}{2}\right]{-}1\right\} & \quad a=2n{+}1,\:n\in\N^{\geq2}. \nonumber
\end{align}

The formulas for $\lambda=(0,a)$ and the formulas for the other half of the points obtained from $\lambda=(a,a)$ are derived by interchanging the coordinates as $(x,y)\rightarrow(y,x)$ of \eqref{H2mult1} and \eqref{H2mult2}, respectively. 

For the Coxeter group $H_3$, the formulas for the coordinates of dominant points of lower orbits are listed in Table~\ref{subtract}. The notation $\big[\cdot\big]$ corresponds to the integer part of a number. In order to generalize each case depending on the type of a dominant point, we only consider $a,b,c\in\{1,2,\dots,9\}$. However, such a generalization can be potentially obtained for any $a,b,c\in\mathbb{N}$. 

In the case of the $H_3$ group, applying reflections given by the Formula \eqref{refl} to dominant points of lower orbits, we obtain the structures of nested polytopes, with their vertices provided in the $\omega-$basis. In Examples~\ref{exm6} and ~\ref{exm7}, to demonstrate the geometric structure of nested polytopes in $\R^3$, the~orthonormal $\alpha-$ and $\omega-$bases of $H_3$ defined in \cite{CMP} are utilized.

The subtraction paths for the non-crystallographic group $H_4$ can be constructed in a similar way. 
However, for $a,b,c,d>0$, an orbit of such a group contains the large number of elements: $|O_{(a,b,c,d)}(H_4)| = 120^{2}$. In this case, the computational routine becomes laborious. Even though for the non-crystallographic cases, the actual method for determining such multiplicities has not yet been developed, it will likely prove related to determining the multiplicities for crystallographic reflection~groups.

In general, to obtain the dominant points of lower orbits, one can choose $\lambda$ with any non-negative real numbers as its coordinates. As shown in Example~\ref{H2tree}, the values from the ring $\mathbb{Z}[\tau]$ can as well represent the coordinates of a dominant point $\lambda$. However, such a choice does not affect the subtraction path. Moreover, it is worth mentioning that the definitions of indices introduced in previous sections of this paper also apply to any lower orbits obtained using the introduced algorithm.

\begin{table}[H]
\caption{Dominant points for lower orbits obtained by subtraction of the simple roots $\alpha_1, \alpha_2, \alpha_3$ of $H_3$ are listed for any type of a dominant point of the initial orbit: $(a,0,0)$, $(0,a,0)$, $(0,0,a)$, $(a,a,0)$, $(0,a,a)$, $(a,0,a)$. The coefficients are provided by the values $a\in\{1,2,\ldots,9\}$.} \label{subtract}
\centering
\def\arraystretch{1.2}
\begin{tabular}{lr}
\hline
\textbf{\boldmath{$(a, 0, 0)$}:} & \\
\hline
$(a-2k, k, 0),\quad k\in\left\{0, \ldots, \left[\frac{a}{2}\right]\right\}$ & \textrm{ any } $a$ \\
$\left(0, \frac{a}{2}(\tau-1), 0\right)$ & \textrm{ even } $a$ \\
$ \left(0, \left[\frac{a}{2}\right]\tau-\left[\frac{a+2}{2}\right], \tau\right)$ &\textrm{ odd } $a>3$ \\
\hline
\textbf{\boldmath{$(0, a, 0)$}:} & \\
\hline
$(k, a-2k, k\tau),\quad k\in\left\{0, \ldots, \left[\frac{a}{2}\right]\right\}$ & \textrm{ any } $a$\\
$(0, 0, 0),\: \left(\frac{a}{2}(\tau{-}1), 0, \frac{a}{2}\right),\: \left(a, \frac{a}{2}(\tau{-}1), 0\right)$ & \textrm{ even } $a$ \\
$ \left(\left[\frac{a}{2}\right]\tau{-}\left[\frac{a{+}2}{2}\right], \tau{+}1, \left[\frac{a}{2}\right]{-}\tau\right),\: \left(a, \left[\frac{a}{2}\right]\tau{-}\left[\frac{a{+}2}{2}\right], \tau\right)$ & \textrm{ odd } $a>3$ \\
\hline
\textbf{\boldmath{$(0, 0, a)$}:} & \\
\hline
$(0, k \tau, a-2k), \quad  k\in\left\{0, \ldots, \left[\frac{a}{2}\right]\right\}$ & \textrm{ any } $a$, \\
$\left(0, \frac{a}{2}(\tau-1), 0\right),\: \left(\frac{a}{2}\tau, 0, \frac{a}{2}(\tau-1)\right)$ & \textrm{ even } $a$ \\
$ \left(\left[\frac{a}{2}\right]\tau, \tau, \left[\frac{a}{2}\right]\tau{-}\left[\frac{a{+}2}{2}\right]\right),\: \left(\tau{+}1, \left[\frac{a}{2}\right]\tau{-}\left[\frac{a{+}2}{2}\right], 0\right)$ & \textrm{ odd } $a>3$ \\
\hline
\textbf{\boldmath{$(a, a, 0)$}:} & \\
\hline
$ (a ,a, 0),\: (0, 0, a \tau),\: (a, a(\tau{-}1), 0)$ &\textrm{ any } $a$ \\
$\left(a-2k, a+k, 0\right),\: (a+k, a-2k, k\tau), \quad  k\in\left\{1, \ldots, \left[\frac{a}{2}\right]\right\}$ & $a>1$  \\
$ \frac{a}{2}\left(2\tau{-}1, 0, 2{-}\tau \right),\: \frac{a}{2}\left(0, \tau{-}1, 0\right), \frac{a}{2}\left(4, \tau-1, 0 \right),\: \frac{a}{2}\left(0, 2-\tau, a\right)$ & \textrm{ even } $a$ \\
$(a, (a-1)\tau-(a+1), 2\tau)$ & $a>4$\\
$\left(2a, \left[\frac{a}{2}\right]\tau-\left[\frac{a}{2}+1\right], \tau\right),\: \left(0, \left[\frac{a}{2}\right]\tau-\left[\frac{a}{2}+1\right], \tau\right)$ & \textrm{ odd } $a>3$ \\
$(a, (a-2)\tau-(a+2), 4\tau)$ & $a>8$\\
\hline
\textbf{\boldmath{$(a, 0, a)$}:} & \\
\hline
$ (a, 0, a), \: (a\tau, 0, 0)$ &\textrm{ any } $a$\\
$(a{-}2k, k, a), \: (a, k\tau, a{-}2k),\quad k\in\left\{1, \ldots, \left[\frac{a}{2}\right]\right\}$&  $a>1$, \\
$\left(0, (a{-}2k{-}1)\tau{-}\left[\frac{a}{2}{+}k{+}1\right], (2k{+}1)(\tau{+}1)\right), \quad k\in\left\{0, \ldots, \left[\frac{a{-}2}{4}\right]\right\}$ & $a>1$ \\
$\frac{a}{2}(0, 1, 0),\: \frac{a}{2}(\tau{+}2, 0, \tau{-}1),\: \frac{a}{2}(1, 0, 2{-}\tau),\: \frac{a}{2}(\tau{-}1, 0, 2\tau{-}1)$ &\textrm{ even } $a$\\
$\left(0, (a{-}2k)\tau{-}\frac{a}{2}{-}k, 2k(\tau{+}1)\right), \quad k\in\left\{0, \ldots, \left[\frac{a}{4}\right]\right\}$ & \\
$\left(\tau{+}2, \left[\frac{a}{2}\right]\tau{-}1, 0\right)$ & \textrm{ odd } $a>1$ \\
$\left(\left[\frac{a}{2}\right]\tau{+}a, \tau, \left[\frac{a}{2}\right]\tau{-}\left[\frac{a}{2}{+}1\right]\right),\: \left(\left[\frac{a}{2}\right]\tau{-}\left[\frac{a}{2}{+}1\right], \tau{+}1, (a{-}1)\tau{-}\left[\frac{a}{2}{+}1\right]\right)$ & \textrm{ odd } $a>3$\\
$\left(2\tau{+}4, \left(\frac{a}{2}{-}1\right)\tau{-}2, 0\right)$ &\textrm{ even } $a>4$\\
$\left(3\tau{+}6, \left[\frac{a}{2}{-}1\right]\tau{-}3, 0\right)$ &\textrm{ odd } $a>5$\\
\hline
\textbf{\boldmath{$(0, a, a)$}:} & \\
\hline
$(0, a, a),\: (a (\tau{+}1), 0, 0)$ & \textrm{ any } $a$ \\
$\left(k, a{-}2k, k\tau{+}a\right),\: \left(0, k\tau{+}a, a{-}2k \right),\quad k\in\left\{1, \ldots, \left[\frac{a}{2}\right]\right\}$& $a>1$ \\
$ \left(0, 0, a \right),\: \frac{a}{2}\left(2\tau{-}1, 0, \tau \right),\: \left(0, \frac{a}{2}(\tau{-}1), 0 \right)$ & \textrm{ even } $a$ \\
$ \left(\left(\frac{a}{2} {-} k\right) (\tau {+} 1), 2k(\tau {+} 1), (a {-} 2 k) \tau - \left[\frac{a}{2} + k\right]\right)$ & \textrm{ even } $a$ \\
$ \left(a, (a {-} 2 k) \tau {-} \frac{a}{2} {-} k, 2 k (\tau + 1)\right), \quad k\in\left\{0, \ldots, \left[\frac{a}{4}\right]\right\}$ & \\
$ \left(\left[\frac{a}{2}{-}k\right](\tau{+}1), (2k+1)(\tau{+}1), (a{-}2k{-}1)\tau-\left[\frac{a}{2}{+}k{+}1\right] \right)$ & \textrm{ odd } $a>1$ \\
$\left(a,(a{-}2k{-}1)\tau-\left[\frac{a}{2}{+}k{+}1\right], (2k+1)(\tau{+}1)\right),\quad k\in\left\{0, \ldots, \left[\frac{a-3}{4}\right]\right\}$ &\\
$\left((a{-}1)\tau{-}\left[\frac{a}{2}{+}1\right], 2\tau{+}1, \left[\frac{a}{2}{-}1\right]\tau{-}1 \right)$ & \textrm{ odd } $a>3$ \\
\hline
\end{tabular}
\end{table}

\newpage\begin{Example}\label{H2tree}
Let us consider the orbit of the non-crystallographic group $H_2$ arising from the dominant point with at least one irrational coordinate, namely $(\tau,1)$. The subtraction of the simple roots $\alpha_1$ and $\alpha_2$ of $H_2$ yields the tree-diagram shown in Figure~\ref{multpH2}. 
\end{Example}

\begin{Example}  
Consider the orbits of $H_3$ with the dominant points $(1,0,0)$ and $(0,0,1)$. Thecoordinates of the orbit-points are obtained from the tree-diagrams provided in Figure~\ref{multpH3}.
\end{Example}

\begin{figure}[H]
\centering
\begin{tabular}{cp{2cm}c}
\resizebox{.1\textwidth}{!}{
\begin{tikzpicture}
\node[block](start) {$(2,0,0)$};
\node[block,below of=start] (c1) {$(0,1,0)$};
\node[below of=c1] (c2) {$(-2,2,0)$};
\node[below of=c2] (c3) {$(-1,0,\tau)$};
\node[below of=c3] (c4) {$(0,-2,2\tau)$};
\node[block,below of=c4] (c5) {$(0,-\tau',0)$};
\node[below of=c5] (c6) {$\dots$};
\path [line] (start) -- node[yshift=0em, xshift=0.8em] {$\hspace{3mm}\a_1$} (c1);
\path [line] (c1) -- node[yshift=0em, xshift=0.8em] {$\hspace{3mm}\a_1$} (c2);
\path [line] (c2) -- node[yshift=0em, xshift=0.8em] {$\hspace{3mm}\a_2$} (c3);
\path [line] (c3) -- node[yshift=0em, xshift=0.8em] {$\hspace{3mm}\a_2$} (c4);
\path [line] (c4) -- node[yshift=0em, xshift=0.8em] {$\hspace{3mm}\tau\a_3$} (c5);
\path [line] (c5) -- node[yshift=0em, xshift=0.8em] {$\hspace{3mm}\tau\a_3$} (c6);
\end{tikzpicture}
} & & 
\includegraphics[scale=0.19]{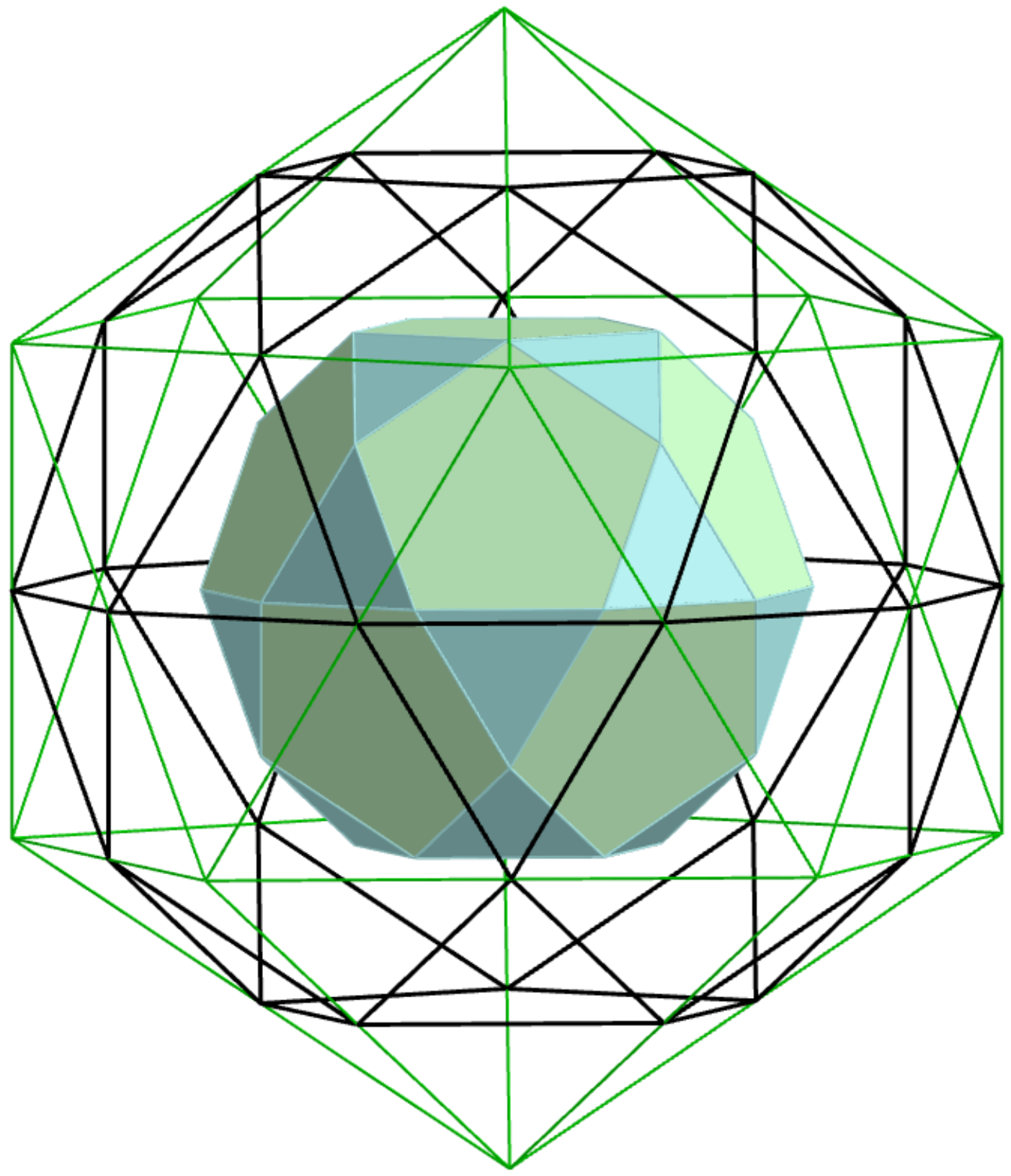}\\
$(\textbf{a})$ & & $(\textbf{b})$\\
\end{tabular}
\caption{(\textbf{a}) The tree-diagram for the orbit $O_{(2,0,0)}(H_3)$; (\textbf{b}) the corresponding nested polytopes. The orbits $O_{(2,0,0)}(H_3)$, $O_{(0,1,0)}(H_3)$ and $O_{(0,-\tau',0)}(H_3)$ are presented in green, black and bold colours,~respectively.}
\label{nestpoly}
\end{figure}
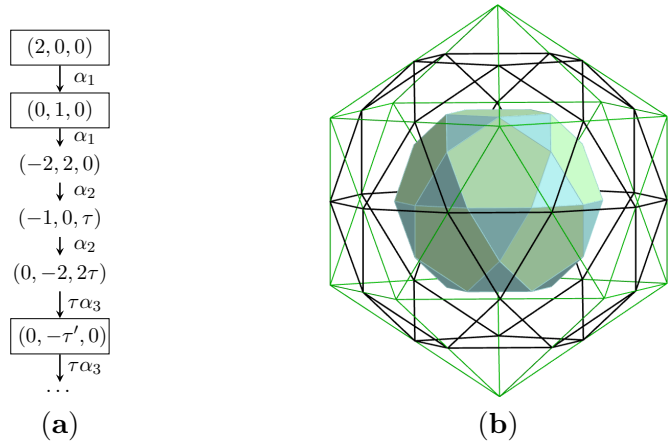

\begin{Example}\label{exm6}
Consider the orbit of $H_3$ with the seed point $(2,0,0)$. As shown in the tree-diagram below, such~an orbit has two lower orbits with the dominant points $(0,1,0)$ and $(0,-\tau',0)$, where $\tau'=1-\tau$. The nested polytopes are generated as presented in Figure~\ref{nestpoly}.
\end{Example}

\begin{figure}[H]
\centering
\begin{tabular}{cp{1cm}c}
\includegraphics[scale=0.17]{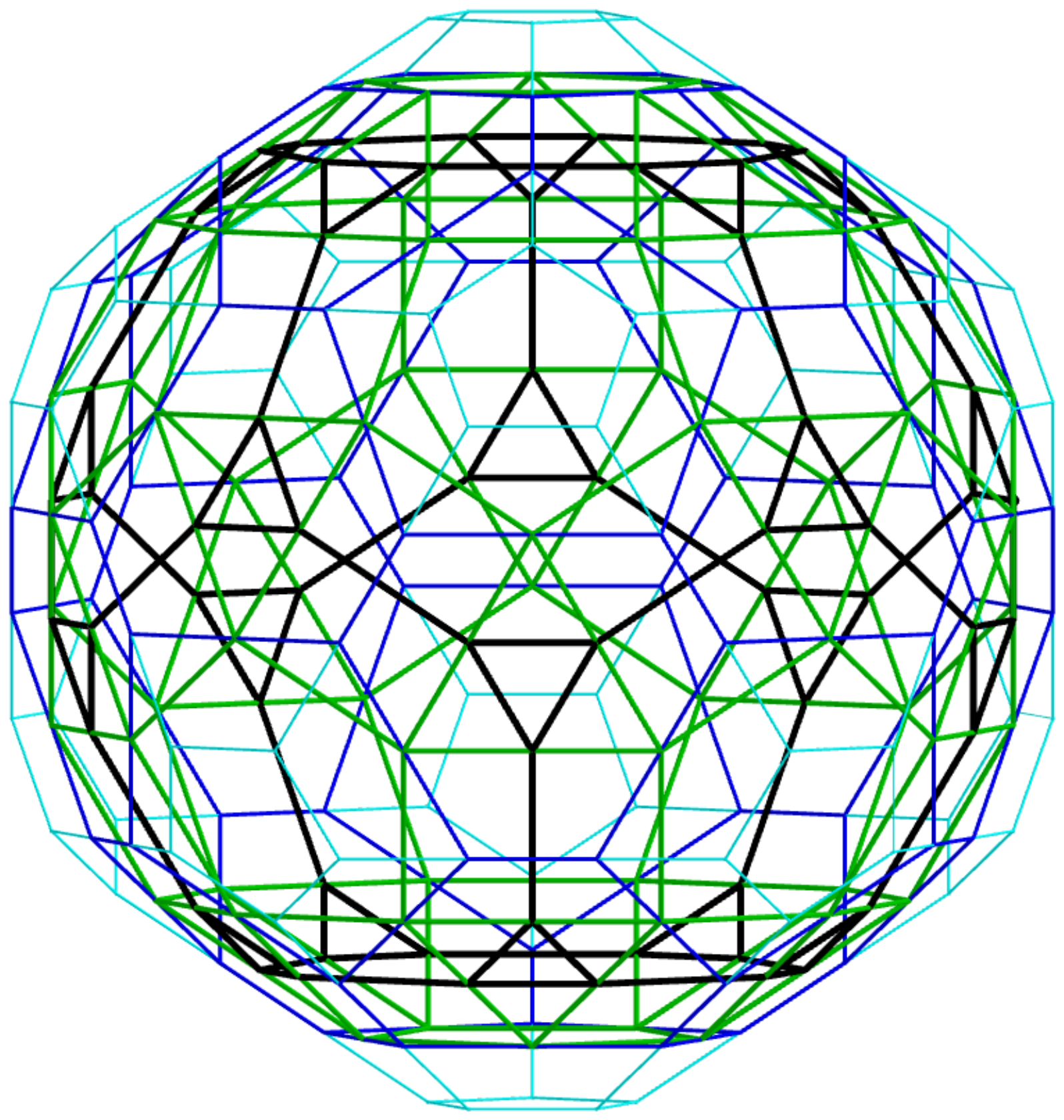} && \includegraphics[scale=0.17]{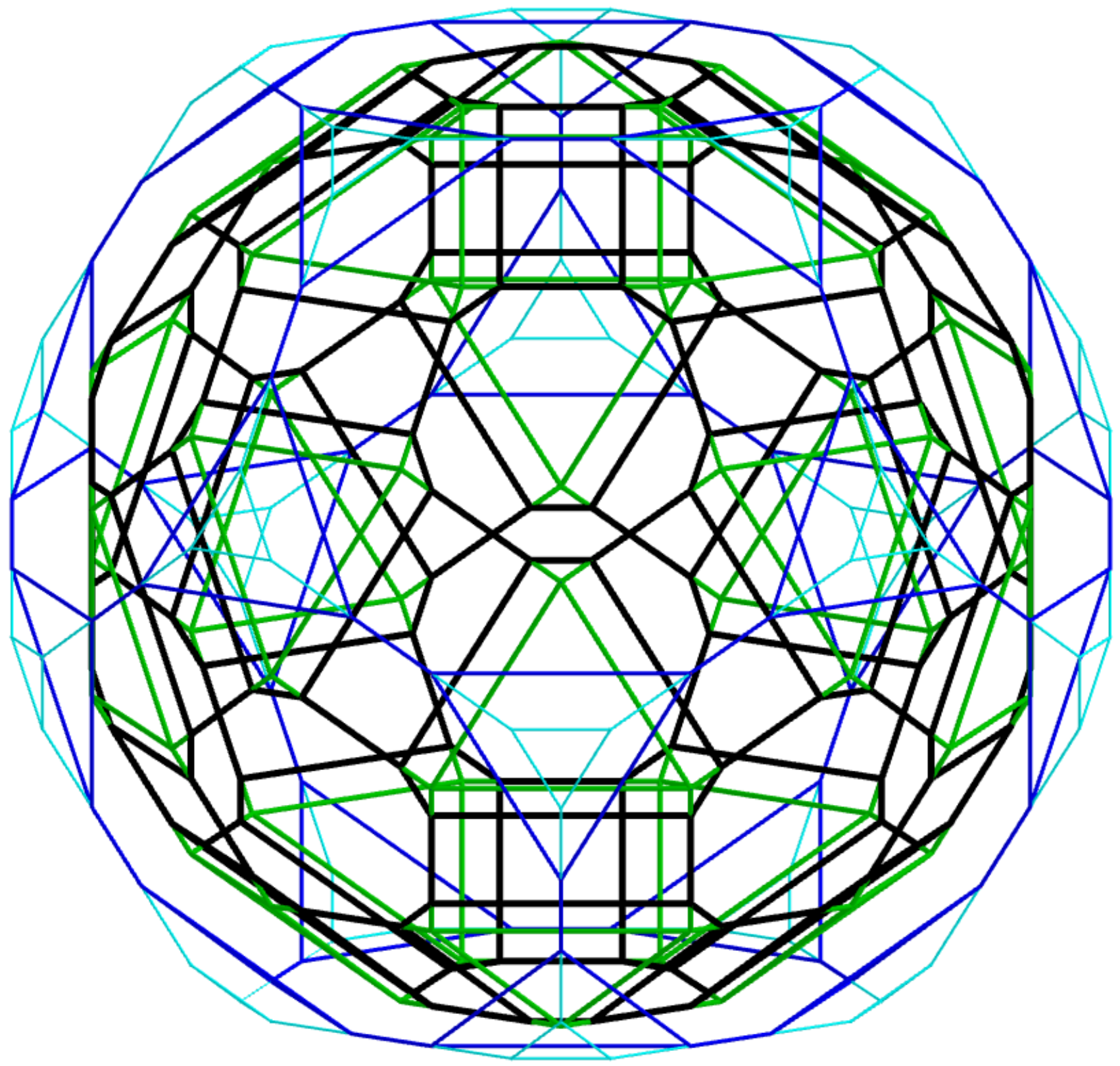} \\
$(\textbf{a})$ && $(\textbf{b})$\\
\end{tabular}
\caption{The nested polytopes provided by the algorithm of root-subtraction. $(\textbf{a})$ $O_{(3,1,0)}(H_3)$; $(\textbf{b})$~$O_{(0,1,3)}(H_3)$.}\label{nest2}
\end{figure}

\begin{Example}\label{exm7}
Consider the orbits $O_{(3,1,0)}(H_3)$ and $O_{(0,1,3)}(H_3)$. Applying the subtraction of the simple roots, we find the following dominant points of lower orbits: 
\begin{small}
\begin{align*}
& (3,1,0): \quad (3,1,0), (1,2,0), (2,0,\tau), (0,1,\tau); \\
& (0,1,3): \quad (0,1,3), (0,\tau+1,1), (\tau+1,0,2), (\tau+1,\tau-1,2\tau-2).
\end{align*}
\end{small}Both of the nested polytopes consist of four orbits of different radii, as shown in Figure~{\ref{nest2}}. Depending on the radius of each orbit that is descending from left to right, they are distinguished by cyan, blue, green and black~colours. 
\end{Example}

\section{Concluding Remarks}

\begin{itemize}

\item The decomposition of a tensor product of representations of a simple Lie algebra into a direct sum of irreducible components given by Young tableaux symmetries plays an essential role in physics. As the indices of the representations help to determine such a decomposition \cite{HLP}, we demonstrate that their definitions can be extended to orbits of the non-crystallographic Coxeter groups. As a result, the notation of the even- and odd-order indices of representations are reformulated for the orbits of $H_n$, $n\in\{2,3,4\}$.
\item It would be useful to generalize the properties of higher-order indices and anomaly numbers of orbits, similarly to \cite{PSW, PS}. Along with these properties, one could potentially obtain the formulas for the explicit forms of higher even-order indices of a tensor product of orbits. Moreover, the expressions for the even-order indices, anomaly numbers and embedding indices could be reformulated and adapted to orbits of any finite reflection group of crystallographic type.

\item Even though the Coxeter groups of non-crystallographic types do not have underlying Lie algebras, the recursive algorithm introduced in Section~\ref{detsec} is shown to be similar to the algorithm developed for the weight multiplicities of simple Lie groups \cite{Brem}. It is important to mention that our algorithm also provides the seed points of orbits that are smaller in radius than an initial orbit (referred to as `lower orbits'). The geometrical construction of sets of lower orbits results in the structures of nested polytopes. Since the recursive rules are only applied to a dominant point $\lambda$ of the non-crystallographic groups $H_2$ and $H_3$, one could consider applying them to any seed point of the $H_4$ group as well. As the size of an orbit $|O_{(a,b,c,d)}(H_4)|=120^2$, for~$a,b,c,d>0$, the~generalization of the formulas for the coordinates of the seed points of lower orbits is considered as future research. Moreover, it would be an interesting task to generalize the formulas given in Table~\ref{subtract} for any $a\in\N$, as it was done for the $H_2$ case.
\end{itemize}



\section*{Acknowledgments}
The authors gratefully acknowledge support from the Natural Sciences and Engineering Research Council of Canada (NSERC), Grant No. RGPIN-2016-04199. The authors are grateful to Dr. A. Patera for helpful comments and editorial assistance.



\begin{thebibliography}{999}

\bibitem{SSP}
Talis, A.L.; Kraposhin, V.S.; Kondrat'ev, S.Y.; Nikolaichik, V.I.; Svyatysheva, E.V.; Everstov, A.A. Non-crystallographic symmetry of liquid metal, flat crystallographic faults and polymorph transformation of the M${\sb 7}$C${\sb 3}$ carbide. {\em Acta Cryst. A} {\bf 2017}, {\em 73}, 209--217. doi:10.1107/S2053273317000936.

\bibitem{NSL}
Nespolo, M.; Souvignier, B.; Litvin, D. About the concept and definition of noncrystallographic symmetry. {\em Z.~Kristallogr. Cryst. Mater.} {\bf 2008}, {\em 223}, 605--606. doi:10.1524/zkri.2008.1137.

\bibitem{Ter}
Terwilliger, T.C. Finding non-crystallographic symmetry in density maps of macromolecular
structures. {\em J.~Struct. Funct. Genom.} {\bf 2013} {\em 14}, 91--95. doi:10.1007/s10969-013-9157-7.

\bibitem{LR}
Levitov, L.S.; Rhyner, J. Crystallography of quasicrystals; application to icosahedral symmetry. {\em J. Phys. France} {\bf 1988}, {\em 49}, 1835--1849. doi:10.1051/jphys:0198800490110183500.

\bibitem{FM}
Fowler, P.W.; Manolopoulos, D.E. {\em An Atlas of Fullerenes}; Dover Publications, Inc.: New York, NY, USA, 2007; ISBN 9780486453620.

\bibitem{Dech}
Dechant, P.-P.; Wardman, J.; Keef, T.; Twarock, R. Viruses and fullerenes---Symmetry as a common thread? {\em Acta Cryst. A} {\bf 2014}, {\em 70}, 162--167. doi:10.1107/S2053273313034220.

\bibitem{IGSRZT}
Indelicato, G.; Cermelli, P.; Salthouse, D.G.; Racca, S.; Zanzotto G.; Twarock, R. A crystallographic approach to structural transitions in icosahedral viruses. {\em J. Math. Biol.} {\bf 2011}, {\em 64}, 745--773. doi:10.1007/s00285-011-0425-5.

\bibitem{SICKT}
Salthouse, D.G.; Indelicato, G.; Cermelli, P.; Keef, T.; Twarock, R. Approximation of virus structure by icosahedral tilings. {\em Acta Cryst. A} {\bf 2015}, {\em 71}, 410--422. doi:10.1107/S2053273315006701.

\bibitem{Tw1}
Twarock, R. Mathematical virology: A novel approach to the structure and assembly of
viruses. {\em Philos. Trans. R. Soc. A} {\bf 2006}, {\em 364}, 3357--3373. doi:10.1098/rsta.2006.1900.

\bibitem{CMP}
Chen, L.; Moody, R.V.; Patera, J. Non-crystallographic root systems. In {\em Quasicrystals and Discrete Geometry}; Patera, J., Ed.; American Mathematical Society: Providence, RI, USA, 1998; Volume 10, pp. 135--178; ISBN~978-0-8218-0682-1.

\bibitem{Shch}
Shcherbak, O.P. Wavefronts and reflection groups. {\em Russ. Math. Surv.} {\bf 1988}, {\em 43}, 149--194. doi:10.1070/rm1988v043n03abeh001741.

\bibitem{CKPS}
Champagne, B.; Kjiri, M.; Patera, J.; Sharp, R.T. Description of reflection-generated polytopes using decorated Coxeter diagrams. {\em Can. J. Phys.} {\bf 1995}, {\em 73}, 566--584. doi:10.1139/p95-084.

\bibitem{Kir}
Kirillov, A., Jr. {\em An Introduction to Lie Groups and Lie Algebras}; Cambridge University Press: Cambridge, UK, 2008; ISBN 9780511755156.

\bibitem{Ram}
Ramond, P. {\em Group Theory: A Physicist's Survey}; Cambridge University Press: Cambridge, UK, 2010; ISBN~9780521896030.

\bibitem{Brad}
Braden, H.W. Integral Pairings and Dynkin Indices. {\em J. London Math. Soc.} {\bf 1991}, {\em s2-43}, 313--323. doi:10.1112/jlms/s2-43.2.313.

\bibitem{Pany}
Panyushev, D.I. On the Dynkin index of a principal $sl_2$-subalgebra. {\em Adv. Math.} {\bf 2009}, {\em 221}, 1115--1121. doi:10.1016/j.aim.2009.01.015.

\bibitem{Dyn}
Dynkin, E.B. Semisimple subalgebras of semisimple Lie algebras. {\em Trans. Am. Math. Soc.} {\bf 1957}, {\em 6}, 111--244. doi:10.1090/trans2/006. 

\bibitem{PSW}
Patera, J.; Sharp, R.T.; Winternitz, P. Higher indices of group representations. {\em J. Math. Phys.} {\bf 1976}, {\em 17}, 1972--1979. doi:10.1063/1.522836.

\bibitem{OP1}
Okubo, S.; Patera, J. General indices of simple Lie algebras and symmetrized product representations. {\em J.~Math. Phys.} {\bf 1983}, {\em 24}, 2722--2733. doi:10.1063/1.525670.

\bibitem{OP2}
Okubo, S.; Patera, J. General indices of representations and Casimir invariants. {\em J. Math. Phys.} {\bf 1984}, {\em 25}, 219--227. doi:10.1063/1.526143.

\bibitem{Okubo1}
Okubo, S. Gauge groups without triangular anomaly. {\em Phys. Rev. D} {\bf 1977}, {\em 16}, 3528--3534. doi:10.1103/PhysRevD.16.3528.

\bibitem{OP3}
Okubo, S.; Patera, J. Cancellation of higher-order anomalies. {\em Phys. Rev. D} {\bf 1985}, {\em 31}, 2669--2671. doi:10.1103/PhysRevD.31.2669.

\bibitem{PS}
Patera, J.; Sharp, R.T. On the triangle anomaly number of $SU(n)$ representations. {\em J. Math. Phys.} {\bf 1981}, {\em 22}, 2352--2356. doi:10.1063/1.524815.

\bibitem{ZOT}
Zhang, H.; Okubo S.; Tosa,Y. Global gauge anomaly for simple Lie algebras. {\em Phys. Rev. D} {\bf 1988}, {\em 37}, 2946--2957. doi:10.1103/physrevd.37.2946.

\bibitem{LNP}
Larouche M.; Nesterenko, M.; Patera J. Branching rules for the Weyl group orbits of the Lie algebra $A_n$. {\em J.~Phys. A} {\bf 2009}, {\em 42}, 485203. doi:10.1088/1751-8113/42/48/485203.

\bibitem{LP2}
Larouche M.; Patera, J. Branching rules for Weyl group orbits of simple Lie algebras $B_n$, $C_n$ and $D_n$. {\em \mbox{J. Phys. A}} {\bf 2011}, {\em 44}, 115203. doi:10.1088/1751-8113/44/11/115203.

\bibitem{GPS}
Grabowiecka, Z.; Patera, J.; Szajewska, M. Reduction of orbits of finite Coxeter groups of non-crystallographic type. {\em J. Math. Phys.} {\bf 2018}, {\em 59}, 101705. doi:10.1063/1.5032210.

\bibitem{Jan}
Janner, A. Alternative approaches to onion-like icosahedral fullerenes. {\em Acta Cryst. A} {\bf 2014}, {\em 70}, 168--180. doi:10.1107/S2053273313034219.

\bibitem{TTVZ}
Thomas, B.G.; Twarock, R.; Valiunas, M.; Zappa, E. Nested Polytopes with Non-crystallographic Symmetry Induced by Projection. In {  Proceedings of the Bridges: Mathematical Connections in Art, Music and Science 2015, Baltimore, MD, USA, 29 July--2 August 2015}; Tessellations Publishing: Phoenix, AZ, USA, 2015; pp.~167--174.

\bibitem{Zel}
Zelevinsky, A. Nested complexes and their polyhedral realizations. {\em Pure Appl. Math. Q.} {\bf 2006}, {\em 2}, 655--671. doi:10.4310/PAMQ.2006.v2.n3.a3.

\bibitem{HLP}
H{\'{a}}kov{\'{a}}, L.; Larouche, M.; Patera, J. The rings of $n$-dimensional polytopes. {\em J. Phys. A} {\bf 2008}, {\em 41}, 495202. doi:10.1088/1751-8113/41/49/495202.

\bibitem{Okubo2}
Okubo, S. Branching index sum rules for simple Lie algebras. {\em J. Math. Phys.} {\bf 1985}, {\em 26}, 2127--2137. doi:10.1063/1.526835.

\bibitem{McKP}
McKay, W.G.; Patera, J. {\em Tables of Dimensions, Indices, and Branching Rules for Representations of Simple Lie Algebras}; Marcel Dekker, Inc.: New York, NY, USA, 1981; ISBN 0824712277.

\bibitem{Brem}
Bremner, M.R.; Moody R.V.; Patera, J. {\em Tables of Dominant Weight Multiplicities for Representations of Simple Lie Algebras}; Marcel Dekker, Inc.: New York, NY, USA, 1985; ISBN 0824772709.
\bibitem{MP82}
Moody, R.V.; Patera, J. Fast recursion formula for weight multiplicities. {\em Bull. Am. Math. Soc.} {\bf 1982}, {\em 7}, 237--242. doi:10.1090/S0273-0979-1982-15021-2.

\bibitem{Brem1}
Bremner, M.R. Fast computation of weight multiplicities. {\em J. Symb. Comput.} {\bf 1986}, {\em 2}, 357--362. doi:10.1016/S0747-7171(86)80003-7.


\end{thebibliography}
\end{document}